\documentclass[useAMS,usenatbib]{mn2e}

\usepackage{graphicx}
\usepackage{amsmath}
\usepackage[dvips,usenames]{color}
\def\aj{AJ}%
\def\apj{ApJ}%
\def\apjl{ApJ}%
\def\apjs{ApJS}%
\def\aap{A\&A}%
\def\aaps{A\&AS}%
\def\mnras{MNRAS}%

\newcommand{\dmo}{\mbox{$(m\!-\!M)_{0}$}}

\newcommand{\av}{\mbox{$A_V$}}
\newcommand{\ebv}{\mbox{$E_{B\!-\!V}$}}

\newcommand{\feh}{\mbox{\rm [{\rm Fe}/{\rm H}]}}

\newcommand{\chisqmin}{\mbox{$\chi^2_{\rm min}$}}

\newcommand{\comment}[1]{}
\newcommand{\beq}{\begin{equation}}
\newcommand{\eeq}{\end{equation}}
\newcommand{\beqa}{\begin{eqnarray}}
\newcommand{\eeqa}{\end{eqnarray}}

\title[The SFH of NGC~1751]
{The star formation history of the Large Magellanic Cloud star cluster
NGC~1751\,\thanks{Based on observations with the NASA/ESA {\it
Hubble Space Telescope}, obtained at the Space Telescope Science
Institute, which is operated by the Association of Universities for
Research in Astronomy, Inc., under NASA contract NAS5-26555}}

\author[Rubele et al.]{Stefano Rubele$^{1,2}$,
	L\'eo Girardi$^{1}$, Vera Kozhurina-Platais$^{3}$, \newauthor
        Paul Goudfrooij$^{3}$, Leandro Kerber$^{4}$  \\
$^{1}$ Osservatorio Astronomico di Padova -- INAF,
	Vicolo dell'Osservatorio 5, I-35122 Padova, Italy \\
$^{2}$ Dipartimento di Astronomia, Universit\`a di Padova,
	Vicolo dell'Osservatorio 2, I-35122 Padova, Italy \\
$^{3}$ Space Telescope Science Institute, San Martin Drive, 
	Baltimore, USA \\
$^{4}$ 
Universidade Estadual de Santa Cruz, 
Rodovia Ilh\'eus-Itabuna, km. 16 -- 
45662-000 Ilh\'eus, Bahia, Brazil
}

\begin{document}

\date{To appear in MNRAS}

\pagerange{\pageref{firstpage}--\pageref{lastpage}} \pubyear{2011}

\maketitle

\label{firstpage}

\begin{abstract}

The HST/ACS colour--magnitude diagrams (CMD) of the populous LMC star
cluster NGC~1751 present both a broad main sequence turn-off and a
dual clump of red giants. We show that the latter feature is real and
associate it to the first appearance of electron-degeneracy in the
H-exhausted cores of the cluster stars. We then apply to the NGC~1751
data the classical method of star formation history (SFH) recovery via
CMD reconstruction, for different radii corresponding to the cluster
centre, the cluster outskirts, and the underlying LMC field. The mean
SFH derived from the LMC field is taken into account during the stage
of SFH-recovery in the cluster regions, in a novel approach which is
shown to significantly improve the quality of the SFH results. For the
cluster centre, we find a best-fitting solution corresponding to
prolonged star formation for a for a timespan of 460~Myr, instead of
the two peaks separated by 200 Myr favoured by a previous work based
on isochrone fitting. Remarkably, our global best-fitting solution
provides an excellent fit to the data -- with $\chi^2$ and residuals
close to the theoretical minimum -- reproducing all the CMD features
including the dual red clump. The results for a larger ring region
around the centre indicate even longer star formation, but in this
case the results are of lower quality, probably because of the
differential extinction detected in the area. Therefore, the presence
of age gradients in NGC1751 could not be probed.  Together with our
previous findings for the SMC cluster NGC~419, the present results for
the NGC~1751 centre argue in favour of multiple star formation
episodes (or continued star formation) being at the origin of the
multiple main sequence turn-offs in Magellanic Cloud clusters with
ages around 1.5~Gyr.

\end{abstract}


\section{Introduction}
\label{intro}

A few stars clusters in the Magellanic Clouds present in their CMDs,
in addition to multiple main sequence turn-offs
\citep[MMSTO;][]{Mackey_BrobyNielsen2007, Mackey_etal08,
Milone_etal08, Goudfrooij_etal09, Glatt_etal08}, also dual red clumps
\citep{Piatti_etal99, Girardi_etal09}. The MMSTO features can be 
interpreted either as the presence of different generations of stars
spanning several $10^8$~yr in these clusters
\citep{Mackey_BrobyNielsen2007, Mackey_etal08, Milone_etal08,
Goudfrooij_etal09}, or as the manifestation of some other effect
intrinsic to coeval stars such as a dispersion in rotational
velocities (\citealt{BastiandeMink09}, see however
\citealt{Girardi_etal11}). However, the simultaneous presence of a
dual red clump feature favours the former interpretation
\citep{Girardi_etal09, Rubele_etal10}: it indicates a modest spread in
the core mass of stars leaving the main sequence which is well
compatible with the age spread of a few $10^8$~yr deduced from the
shape of the MMSTOs.

The SMC star cluster NGC~419 is presently the most striking example of
a cluster containing a dual red clump. \citet{Rubele_etal10}
demonstrated that the assumption of an extended star formation history
(SFH) in NGC~419, explored by means of the classical method of
SFH-recovery via CMD-reconstruction, produces indeed a remarkably good
quantitative description of the observed CMDs. The SFH was found to
extend over a period of 700~Myr. The same analysis has produced quite
stringent limits to the cluster's distance, reddening, and
metallicity.

Although not explicitly discussed by \cite{Rubele_etal10}, the dual
red clump of NGC~419 has played an important role in limiting the
family of stellar models that could be fit in the process of
CMD-reconstruction, because dual red clumps can only happen within a
relatively narrow interval of ages.
In this paper, we examine the case of the LMC star cluster NGC~1751,
which as noted by \citet{Girardi_etal09}, does also appear to present
a dual red clump, and hence should be a good target for the
CMD-reconstruction technique. We will use the extremely accurate data
available from HST/ACS, and analyses techniques similar to those
applied by \citet{Girardi_etal09} and \citet{Rubele_etal10} for
deriving the SFH. Sect~\ref{data} will briefly present the data and
discuss the reality of the dual red clump. The next sections will
present the SFH-recovery method and its application to the NGC~1751
surrounding LMC field (Sect.~\ref{sec_fieldSFH}) and cluster area
(Sect.~\ref{sec_clusterSFH}). Sect.~\ref{conclu} draws the final
conclusions.

\section{The NGC~1751 data and its dual red clump}
\label{data}

\subsection{Data and photometry}
\label{dataphot}

The dataset used in this paper comes from GO-10595 (PI: Goudfrooij),
and consists of one short and two long exposures in F435W, F814W, and
F555W with small dither pattern to avoid the ACS/WFC gap between two
WFC chips. A detailed description of the observations and photometry
is given in \citet{Goudfrooij_etal09}.  Nevertheless, in this paper we
use the simultaneous ePSF fitting technique as it described in
Anderson et.al. (2008), which fits the PSF simultaneously on all
exposures/observations of the cluster.  Differently from
\citet{Goudfrooij_etal09}, the Charge Transfer Efficiency (CTE)
correction was performed using \citet{RiessMack04} formula (ACS-ISR
2005). The derived photometry was calibrated into the Vegamag system
as described in \citet{Goudfrooij_etal09}.

\begin{figure}
\resizebox{\hsize}{!}{\includegraphics{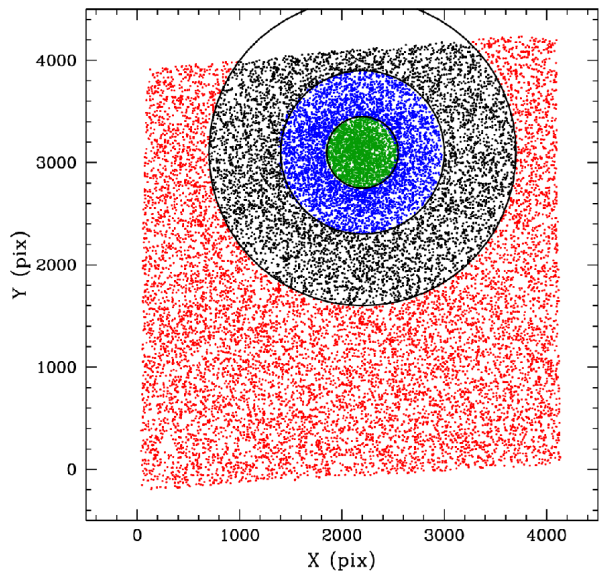}}
\caption{Map of the stars used in this work, in the $xy$ plane of the 
ACS/WFC images. The scale is of about $0.05$~arcsec/pix. The observed
stars have been grouped in areas corresponding to the LMC field (red)
and, for NGC~1751, an inner ``Centre'' (green) and outer ``Ring''
(blue).}
\label{fig_areas}
\end{figure}

\begin{figure}
\resizebox{\hsize}{!}{\includegraphics{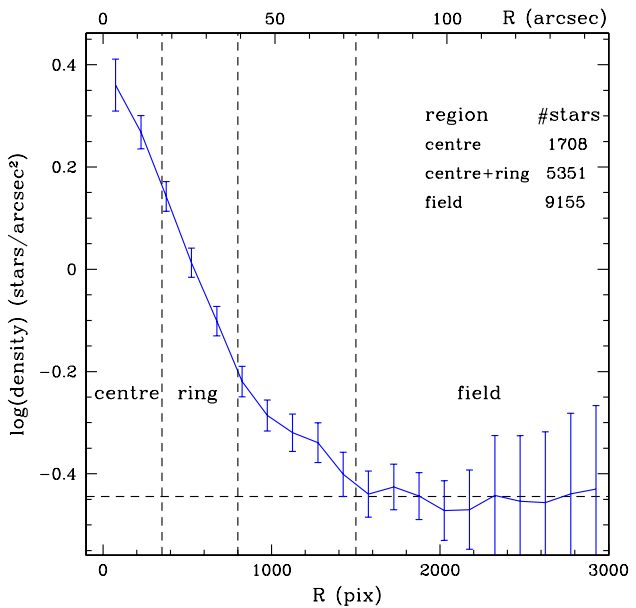}}
\caption{The logarithm of stellar density as a function of radius 
from the NGC~1751 centre. Error bars are the random errors.}
\label{fig_densityprofile}
\end{figure}

Fig.~\ref{fig_areas} shows a spatial representation of the stars we
analize in this work. 
We have initially located the center of NGC~1751 at the position
$x_{\rm c}=2200$~pix, $y_{\rm c}=3100$~pix. Based on the appearance of
the CMDs at varying radii $r$ from this center, we have defined
regions of the ACS/WFC images corresponding to
\begin{itemize}
\item the LMC {\bf Field}, for $r>1500$~pix (73.5~arcsec);
\item the main body of the NGC~1751 cluster, for $r<800$~pix 
(39.2~arcsec), which is further divided into:
\begin{itemize}
\item the {\bf Centre}, for $r<350$~pix (17.2~arcsec);
\item the {\bf Ring}, for $350<r<800$~pix ($17.2<r<39.2$~arcsec).
\end{itemize}
\end{itemize}
These regions are depicted in Fig.~\ref{fig_areas}. The Centre, Ring
and Field regions have areas of 0.385, 1.63, and 10.30 pix$^2$ (5.3,
22.6, 143 arcmin$^2$), respectively.  Figure~\ref{fig_densityprofile}
shows how the stellar density varies as a function of radius from the
NGC~1751 centre, taking into account only the stars of $F814W<22$, for
which the photometry should be close to complete. The figure clearly
shows the flattening of the density for $r>1500$~pix, which indicates
that indeed that is a good choice for defining the LMC Field.

In this work, we will analyse both the Centre and Ring regions, which
present a good density contrast with respect to the field. Although
the cluster clearly extends up to a radius of $1500$~pix, the region
with $800<r<1500$~pix will not be considered further.

\begin{figure}
\resizebox{\hsize}{!}{\includegraphics{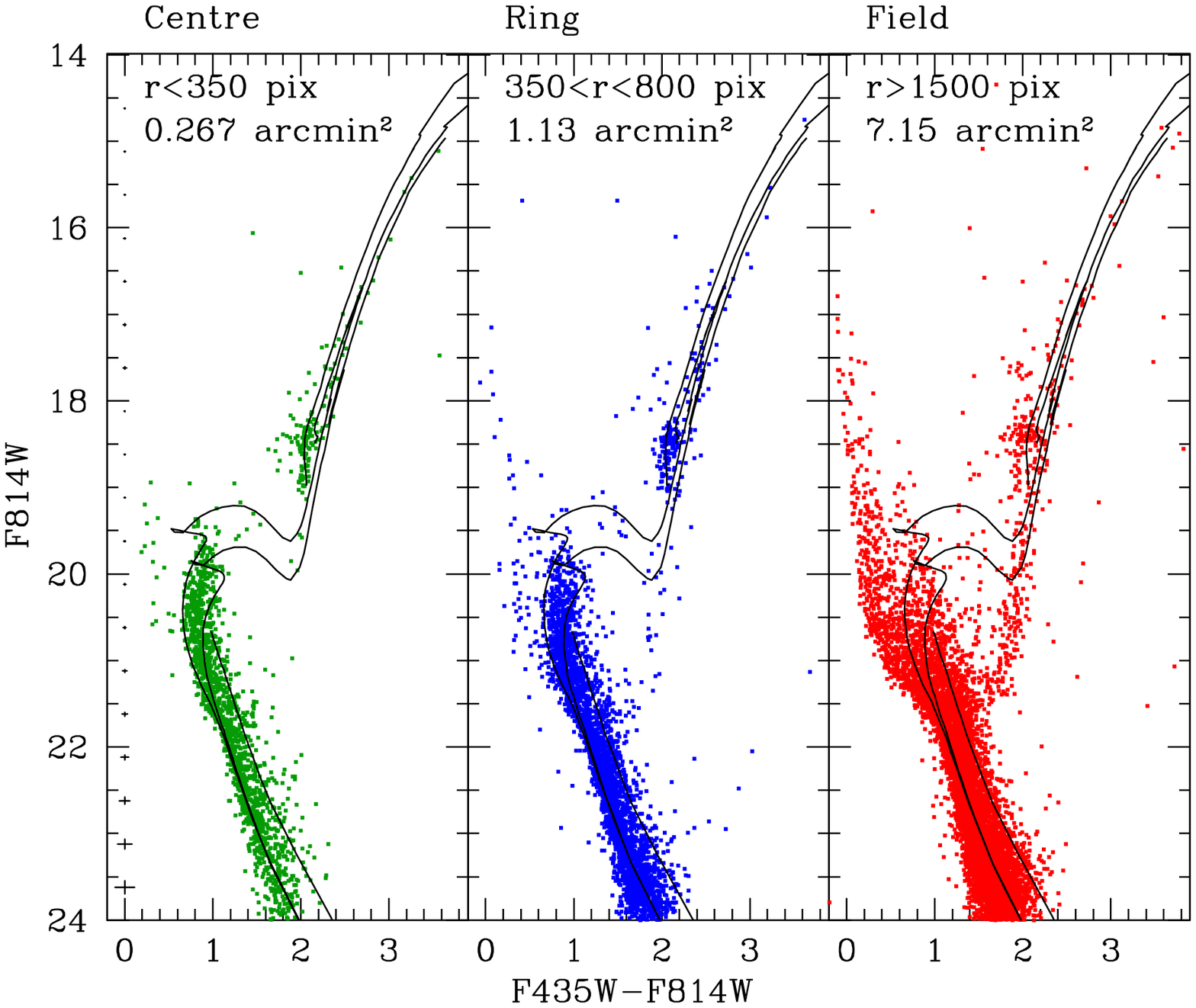}}
\resizebox{\hsize}{!}{\includegraphics{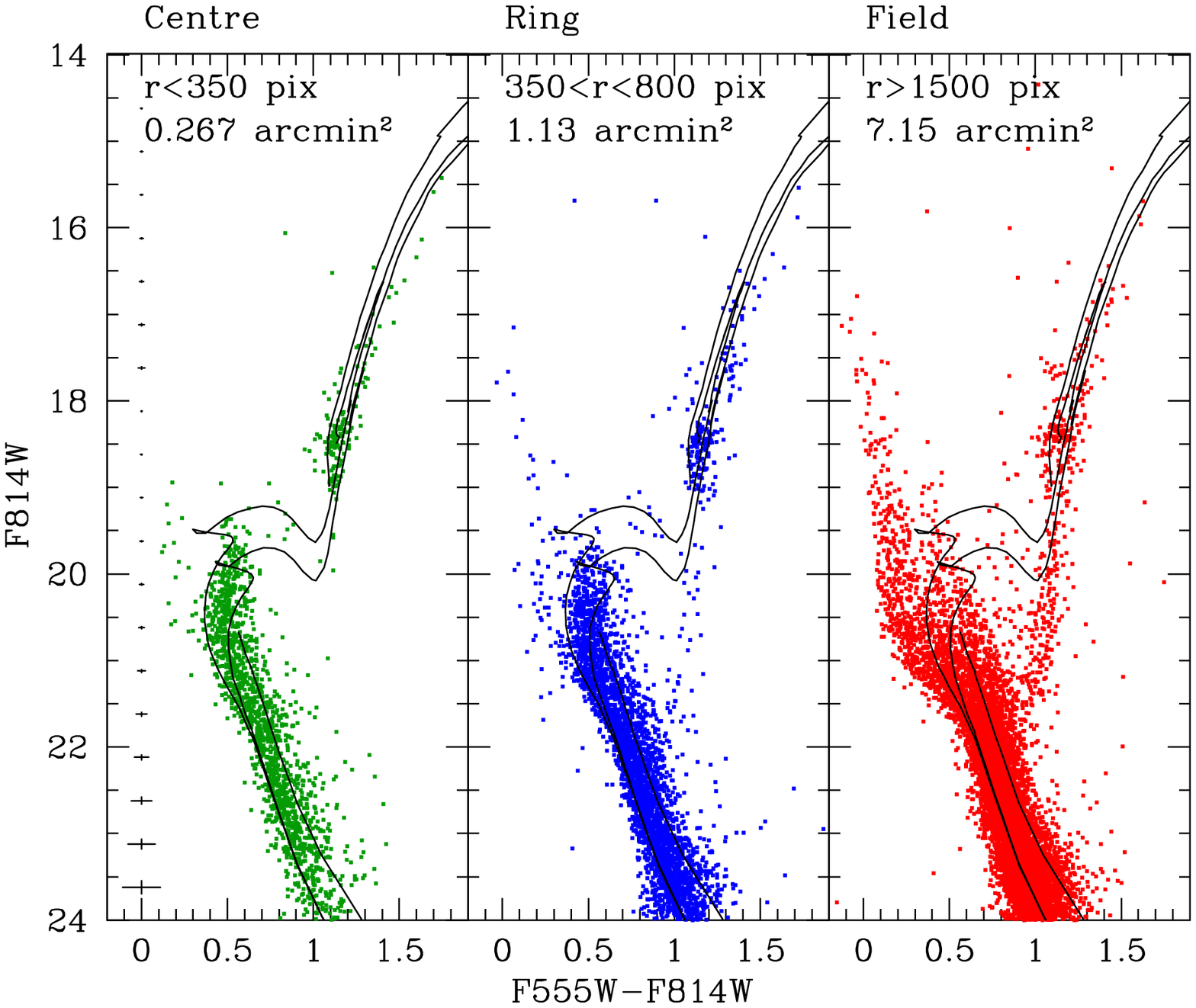}}
\caption{The CMDs for NGC~1751 as derived from the ACS/WFC data, 
using both $F435W-F814W$ (top panels) and $F435W-F814W$ (bottom
panels) colours versus the $F814W$ magnitude. Panels from left to right
present data for the cluster Centre and Ring, and LMC Field. As a
reference to the eye, the panels also show the position of 1.12 and
1.42-Gyr isochrones of metal content $Z=0.008$, shifted by
$\dmo=18.50$ and $A_V=0.7$, together with the expected location of
equal-mass binaries along the main sequence (continuous lines). The
tiny crosses at the leftmost extreme of the left panel are $1\sigma$
error bars derived from artificial star tests in the cluster Centre
(see Sect.~\ref{sec_ast}). The error bars for the Ring and Field are
not shown in the figure; they are of about the same size for the
brightest magnitudes, becoming just $\sim25$~\% smaller for the
faintest magnitudes. }
\label{fig_cmd}
\end{figure}

Fig.~\ref{fig_cmd} shows the ACS data for the different regions of
NGC~1751, in the F814W vs.\ F435W\,$-$\,F814W and F814W vs.\
F555W\,$-$\,F814W CMDs. These plots will be used as a reference in
our analysis.

The CMDs for the cluster Centre show very clearly the broad main
sequence turn-off of this cluster, its dual red clump, and other
well-known CMD features such as the sequence of binaries parallel to
the main sequence, and the RGB, subgiants, and early-AGB bump. A
simple comparison between the CMDs for the Centre and Field reveals
that the field contamination in the cluster central regions is close
to negligible. This is clear already looking at the star counts in the
red clump: the 7.15~arcmin$^2$ of Field contain 189 red clump stars
(here defined as stars with $18.05\!<F814W\!<19.15$,
$F435W\!-\!F814W\!>\!1.5$), therefore the 0.267~arcmin$^2$ of the
Centre are expected to contain just $\sim7$ red clump stars coming
from the LMC field, which is far less than necessary to explain any of
the features of its CMD. Indeed, the red clump in the Centre is made
of 117 stars, which can be separated into the 89 ``bright'' ones
($F814W<18.75$) -- which correspond to the classical red clump made of
stars which likely passed through electron degeneracy in their cores
-- and 28 faint or ``secondary'' ones ($F814W>18.75$) -- which were
likely able to avoid it. We can conclude, in a way similar to the case
of NGC~419 \citep{Girardi_etal09, Girardi_etal09rio}, that the
probability that the dual red clump observed in the centre of NGC~1751
is caused by LMC field stars is less than $P\sim5\times10^{-6}$, and
therefore negligible.

Note that differences are quite evident in the position of CMD
features between the Centre and Field, which are obviously mixed in
the CMD of the Ring. The Field presents an old main sequence turn-off
and subgiant sequence extending to magnitudes as faint as
$F814W<21.5$, and a younger main sequence extending as bright as
$F814W<16.5$. Just traces of these features are present in the CMD of
the Centre. Moreover, the red clump in the Field do also present a
composite structure, with a ratio of faint/bright stars of
59/130. This latter feature is just expected from a field made of
stars covering a wide range of ages and initial masses
\citep[see][]{Girardi99, Piatti_etal99}, and being observed with 
very small photometric errors as in this case.

\subsection{Assessing photometric errors and completeness}
\label{sec_ast}

In order to characterize the errors in the photometry and the
completeness of the sample, we have performed a series of artificial
star tests (AST) on the reduced images \citep[see
e.g.][]{Gallart_etal99,HZ01}. 

The procedure consists of adding stars of known magnitude and colour
at random places in each exposure, and redoing the photometry exactly
in the same way as described in Sect.~\ref{dataphot}. The artificial stars
are considered to be recovered if the input and output positions are
closer than 0.5 pixels, and flux differences are less than 0.5~mag.

In order to avoid the introduction of additional crowding in the
images, artificial stars are positioned at distances much higher than
their PSF width.  So, our AST are distributed on a grid spaced by
20~pix, which is each time randomly displaced over each set of
exposures.

\begin{figure}
\resizebox{0.48\hsize}{!}{\includegraphics{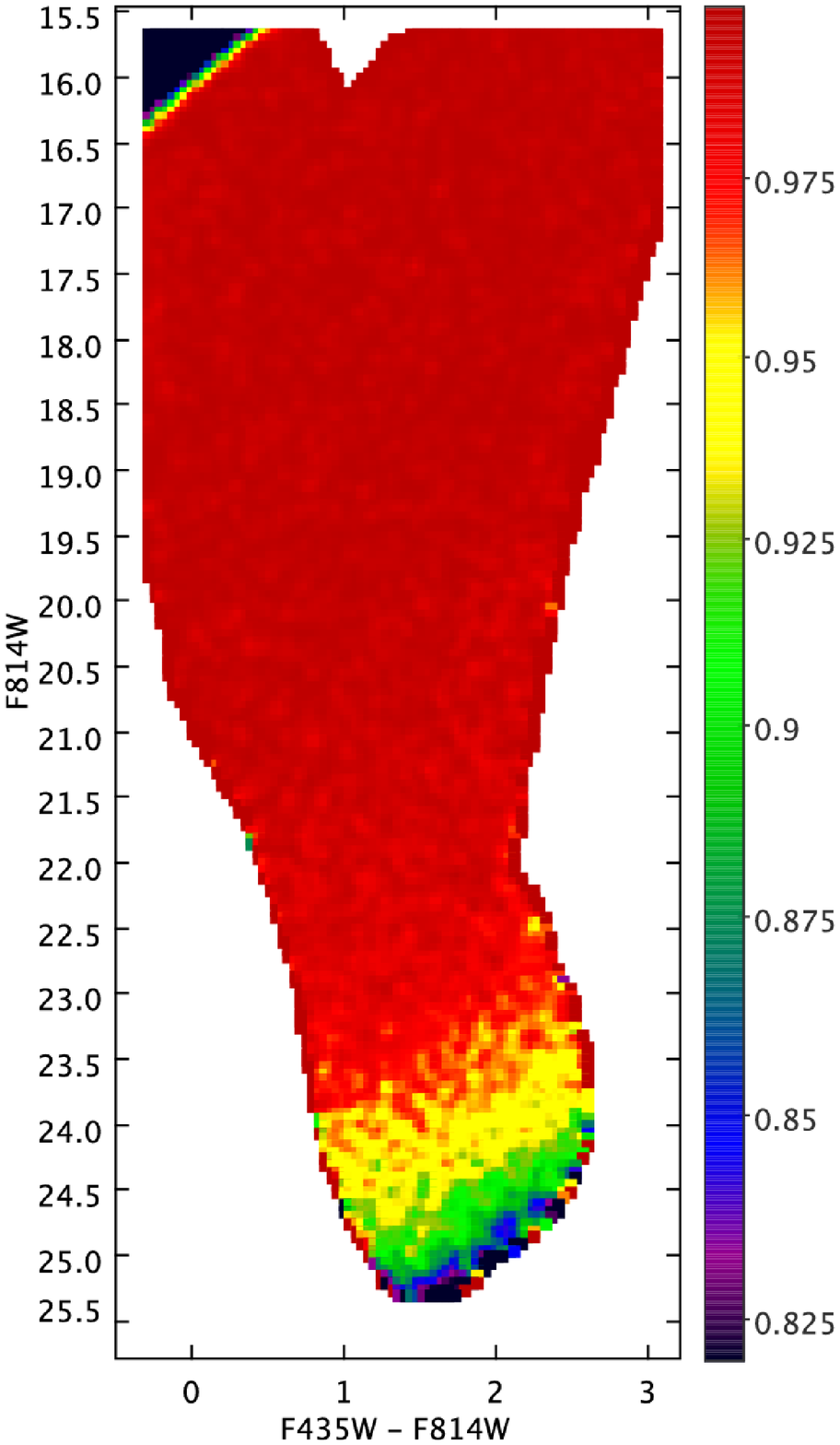}}
\resizebox{0.48\hsize}{!}{\includegraphics{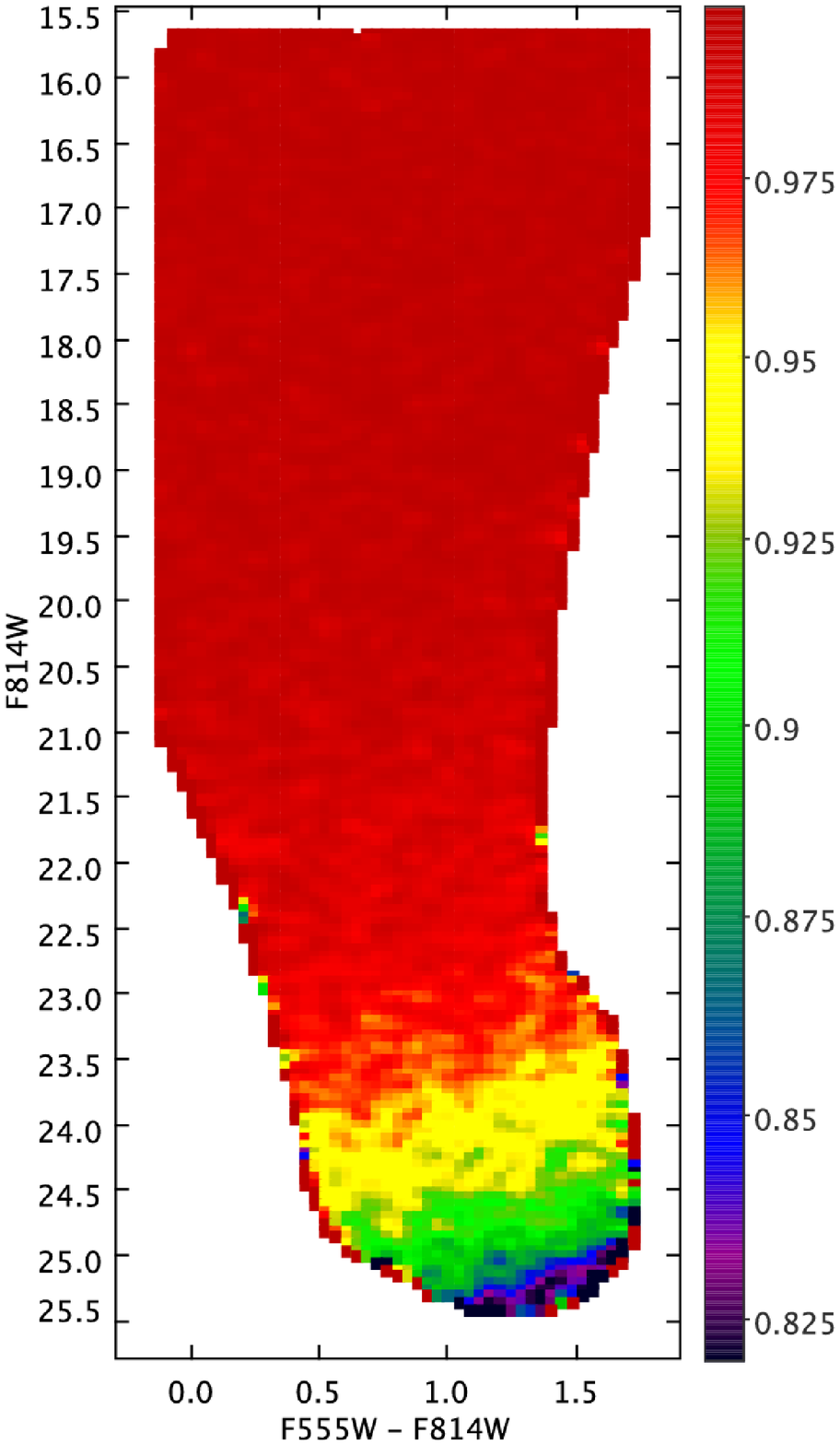}}
\caption{Completeness map, derived from the complete set of ASTs 
realised over NGC~1751 (centre plus ring areas), for both the F814W
vs. F435W$\,-\,$F814W (left panel) and F814W vs. F555W$\,-\,$F814W
(right panel).}
\label{fig_completeness}
\end{figure}

A total of $1.04\times10^7$ ASTs were performed, covering in an almost
uniform way the CMD area of the observed stars as well as the area for
which we build the ``partial models'' to be used in the SFH analysis
(see Sect.~\ref{sec_partialmodels} below). Then, the ratio between
recovered and input stars gives origin to the completeness map of
Fig.~\ref{fig_completeness}. Note that the 90~\% completeness limit is
located at F814W\,$\sim24.5$, which is well below the position of the
MMSTOs in NGC~1751.

\begin{figure}
\resizebox{\hsize}{!}{\includegraphics{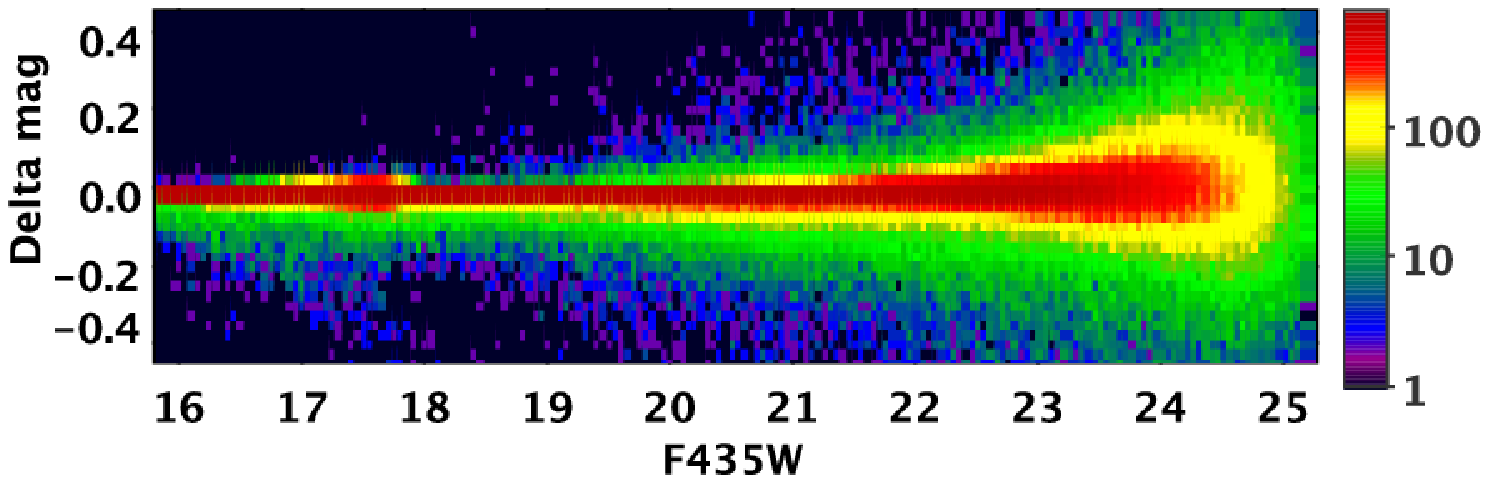}}
\resizebox{\hsize}{!}{\includegraphics{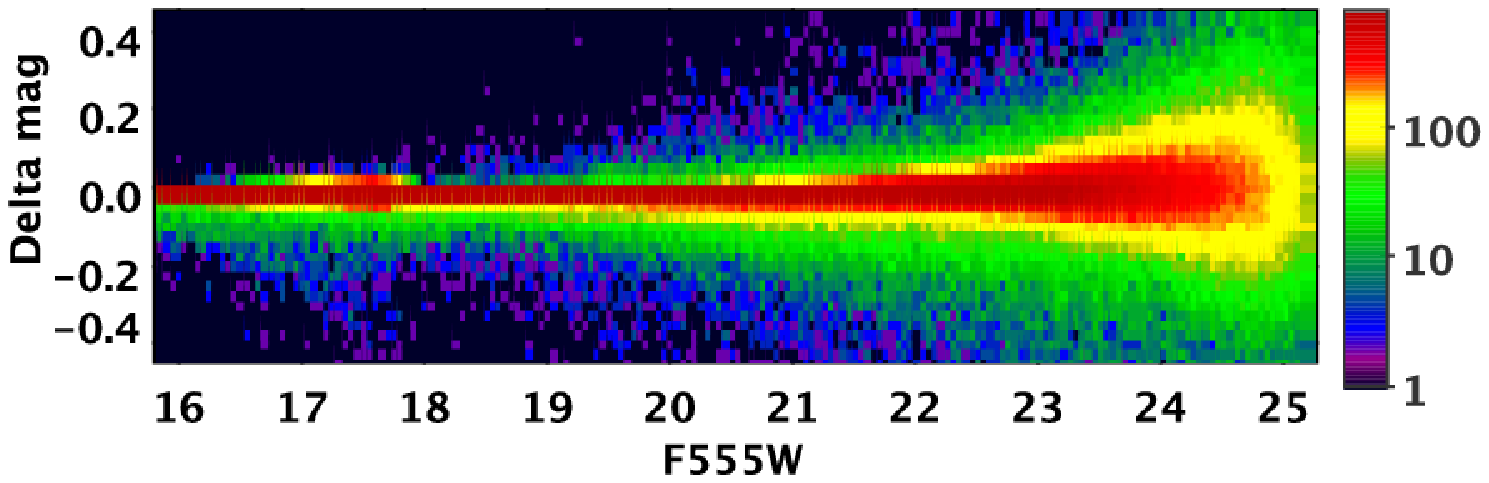}}
\resizebox{\hsize}{!}{\includegraphics{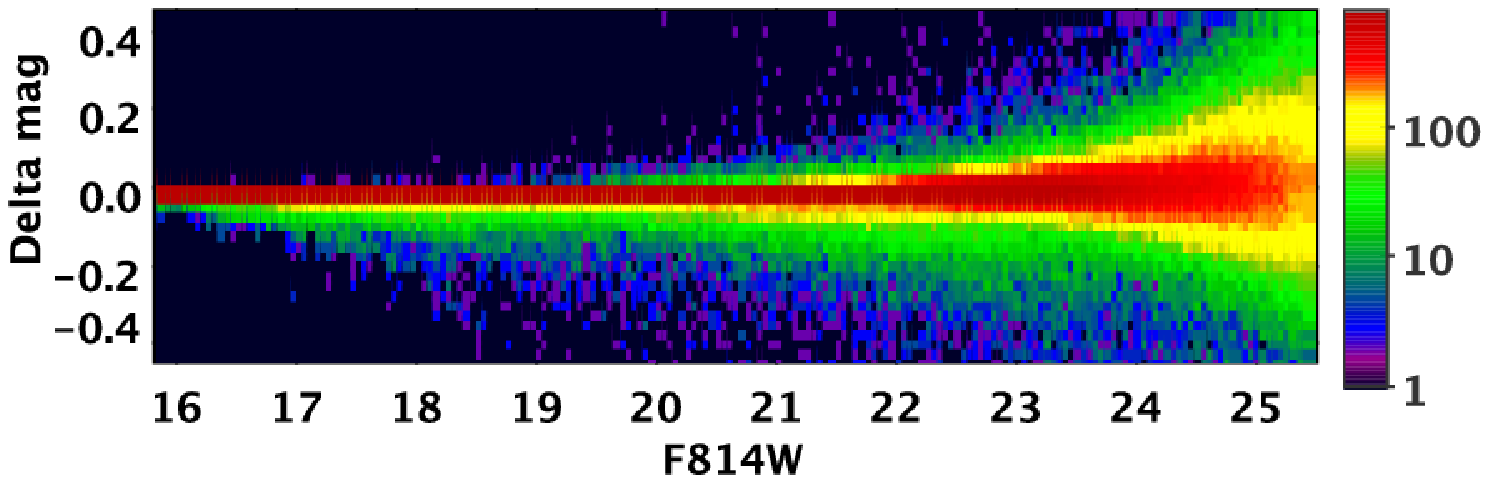}}
\caption{Map of photometric errors as a function of input F435W, F555W and
F814W (from top to bottom), as derived from the ASTs over the core of
the cluster area (that is, in the Centre plus Ring). The errors are
defined as the difference between the recovered and input magnitudes.
}
\label{fig_photerrors}
\end{figure}

Figure~\ref{fig_photerrors} illustrates the differences between the
recovered and input magnitudes of the ASTs, as a function of their
input magnitudes. These differences give a good handle of the
photometric errors effectively present in the data. The error
distributions are slightly asymmetric because of crowding.

\section{The SFH of the LMC Field}
\label{sec_fieldSFH}

\begin{figure}
\resizebox{\hsize}{!}{\includegraphics{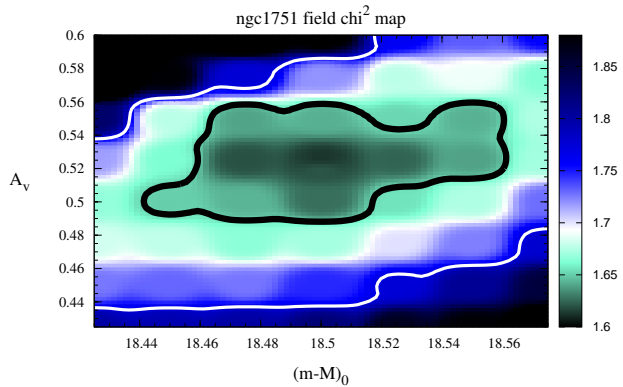}}
\caption{$\chi^{2}$ map for the Field best-fitting solution, as a 
function of distance modulus and $V$-band extinction. The continuous
lines show the $68~\%$ (black) and 95~\% (white) confidence levels for
the overall best-fitting solution, which is located at $\dmo=18.50,
\av=0.525$.}
\label{fig_fieldchimap}
\end{figure}

\subsection{Overview of the method}
\label{sec_overview_recovery}

To recover the SFH from the ACS data, we use the pipeline built to
analyse data from the VISTA survey of the Magellanic Clouds
\citep[VMC; see][]{Cioni_etal08}. The method has been fully described
and tested by \citet{Kerber_etal09} using simulated near-infrared
data, and by \citet{Rubele_etal10} using ACS/HRC data for the SMC
cluster NGC~419. The method consists in (1) building the Hess diagram
for the data and a series of ``partial models'' which represent
populations in limited intervals of age and metallicity, and (2) using
the StarFISH code \citep{HZ01,HZ04} to find the linear combination of
partial models that minimizes a $\chi^2$-like statistic as defined in
\citet{Dolphin02}. The solution is characterized by the minimum
$\chi^2$, $\chisqmin$, and by a set of partial model coefficients
corresponding to the several age bins. These latter translate directly
into the star formation rate as function of age, SFR$(t)$.

The age--metallicity space occupied by the partial models depends on
the object under consideration. In the present work, we have to
consider two distinct cases, corresponding to the cluster and LMC
field as previously defined. This section deals with the field only.

The LMC field is expected to follow a marked age-metallicity relation
(AMR). This AMR has been measured by several authors using both field
stars and star clusters of several ages \citep{MG03, Kerber_etal07,
Grocholski_etal06, Grocholski_etal07}. In addition to the mean AMR, it
is reasonable to expect a modest spread in metallicity at any given
age. For this work, we adopt the scheme set by \citet{Kerber_etal09},
in which we build partial models at 5 different metallicities disposed
around the mean AMR of the LMC. Each partial model covers a range in
logarithm of age of width 0.2~dex. For stellar populations younger
than $10^8$~yr, the numbers of observed stars are very small and hence
we assume broader age bins, of widths 0.3~dex for $7.2\le\log(t/ {\rm
yr})\le8.0$, together with a single age bin of width 0.8~dex for
$\log(t/{\rm yr})<7.2$. The [Fe/H] separation between partial models
is of 0.1~dex.

For the initial mass function (IMF) we adopt the \citet{chabrier01}
one.  The binary fraction is set to a value of 0.3 for binaries with
mass ratios in the range between 0.7 and 1.0, which is consistent with
the prescriptions for binaries commonly used in works devoted to
recover the field SFH in the Magellanic Clouds
\citep[e.g.][]{Holtzman_etal99, HZ01, Javiel_etal05, Noel_etal09}.
Notice that this assumption is also in agreement with the few
determinations of binary fraction for stellar clusters in the LMC
\citep[][both for NGC 1818, a stellar
cluster younger than $\sim100$~Myr]{Elson_etal98, Hu_etal08}.

\begin{figure*}
\resizebox{0.25\hsize}{!}{\includegraphics{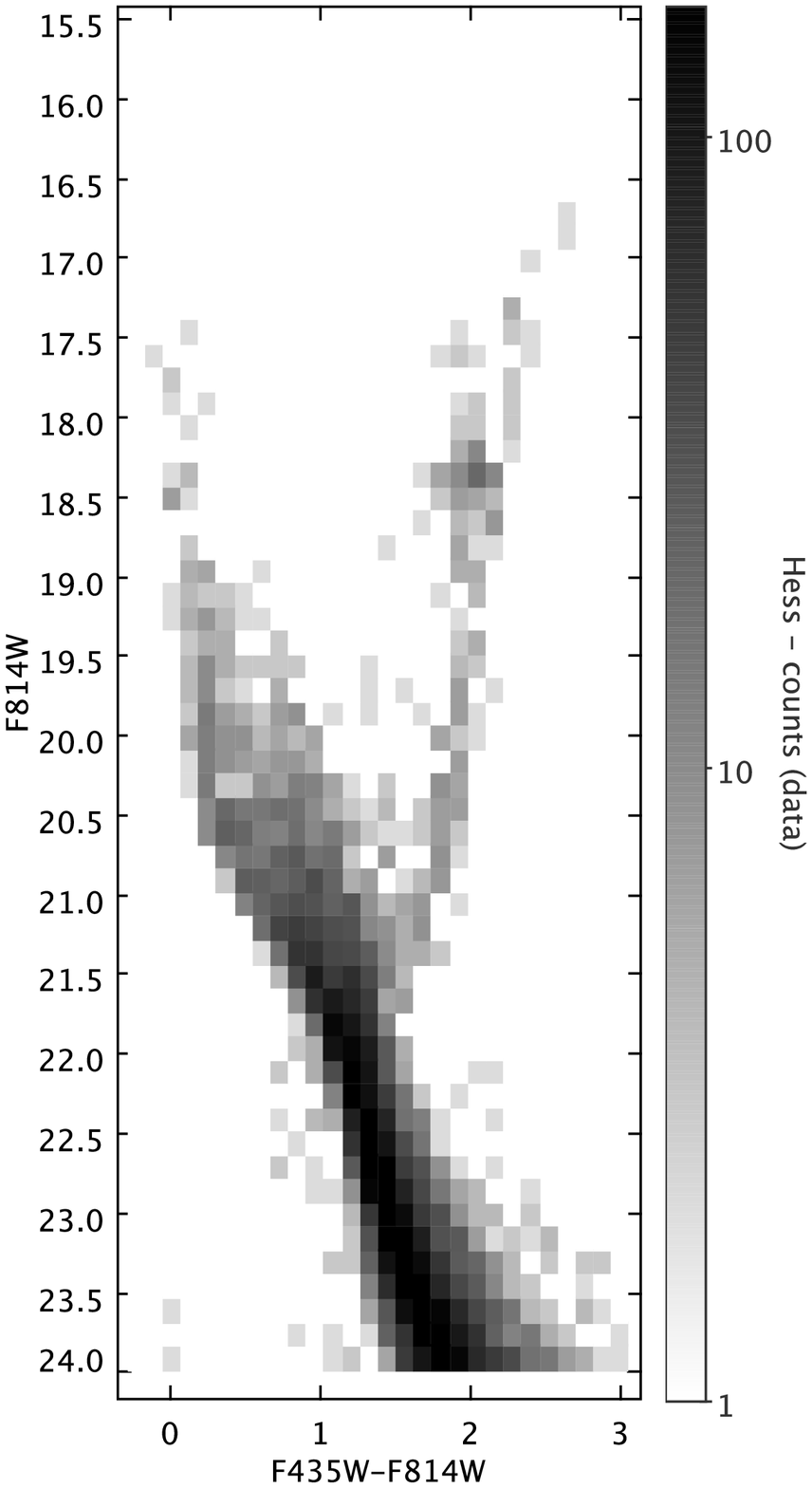}}
\resizebox{0.25\hsize}{!}{\includegraphics{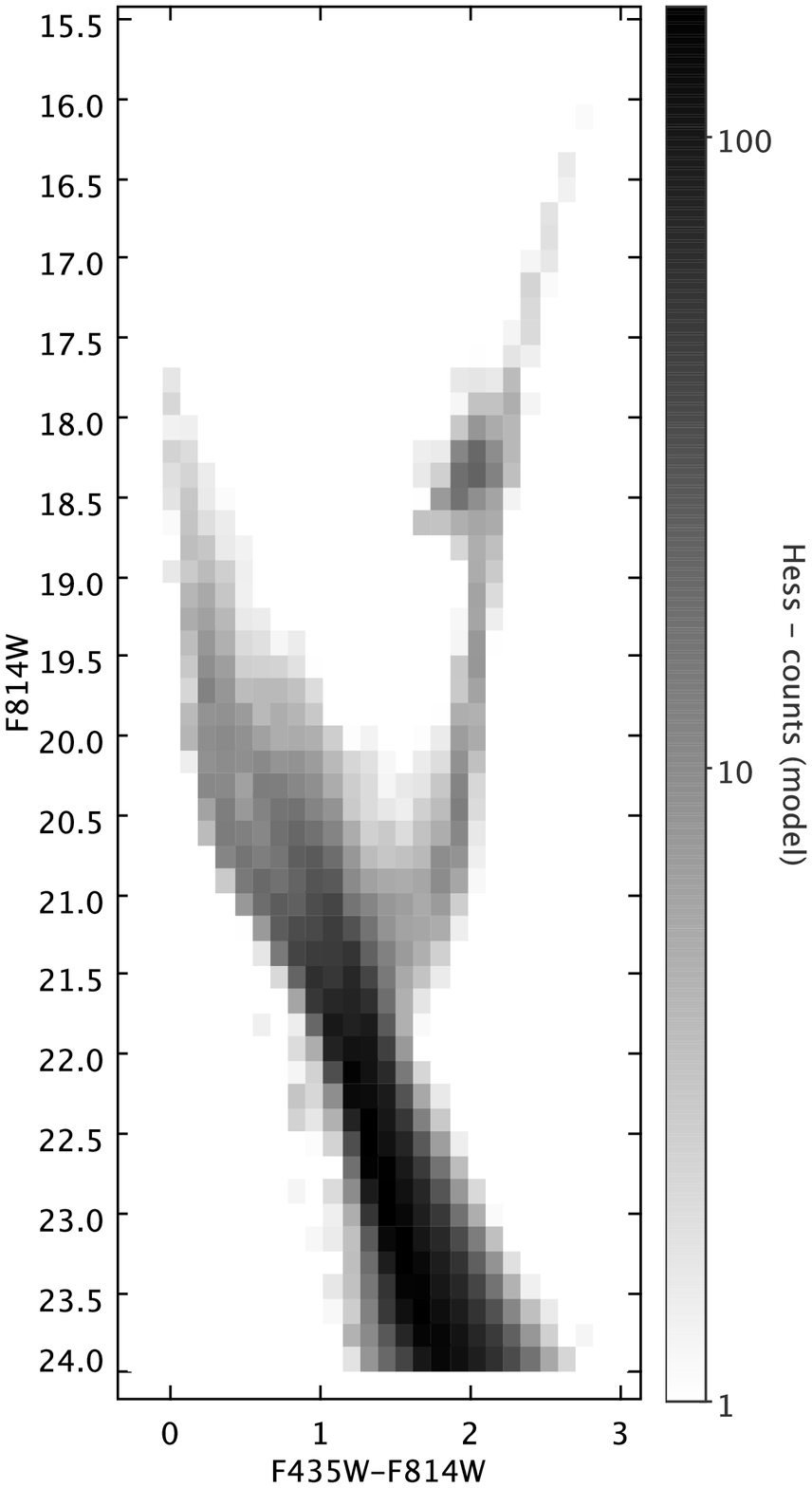}}
\resizebox{0.25\hsize}{!}{\includegraphics{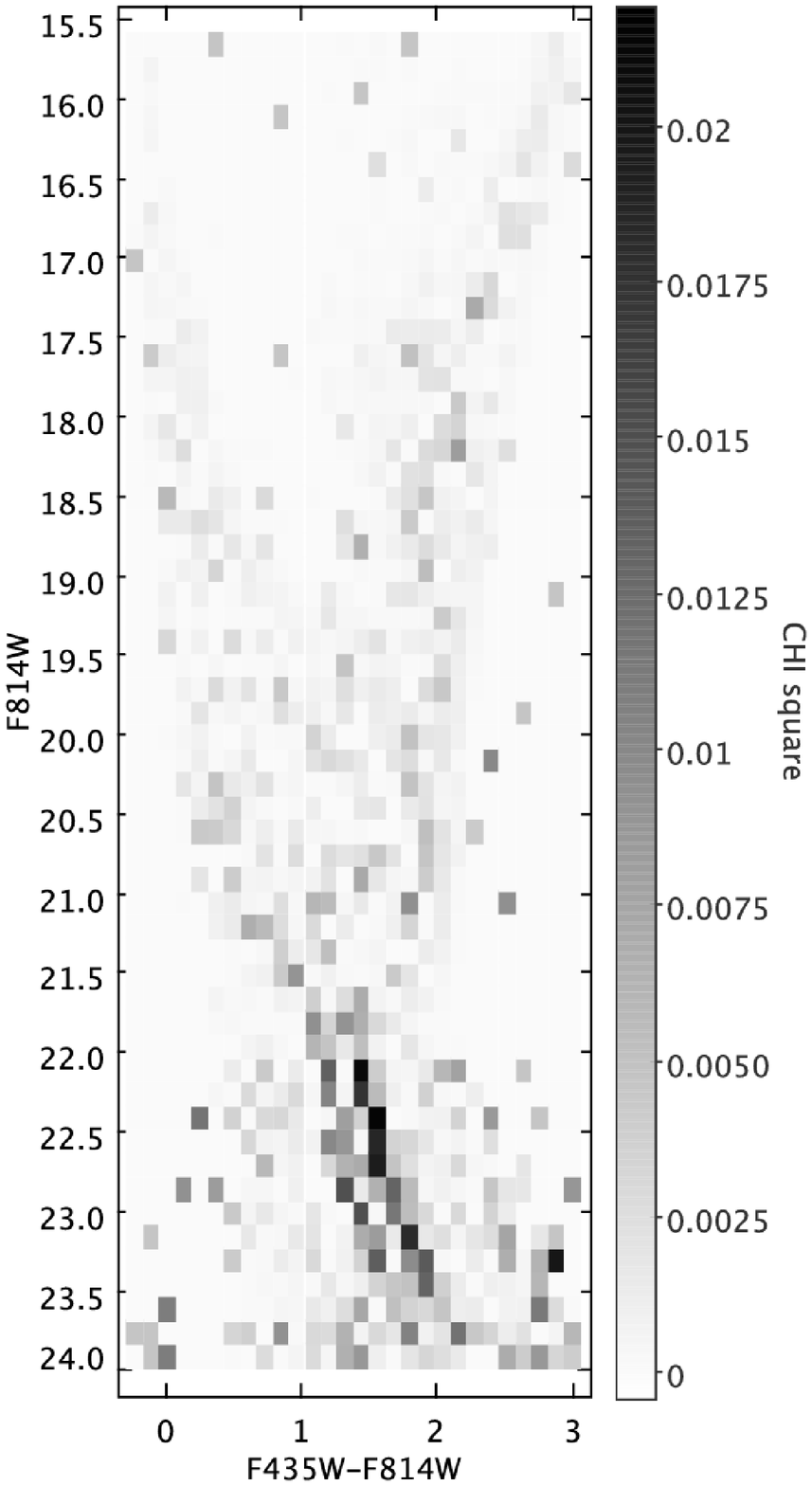}}
\\
\resizebox{0.25\hsize}{!}{\includegraphics{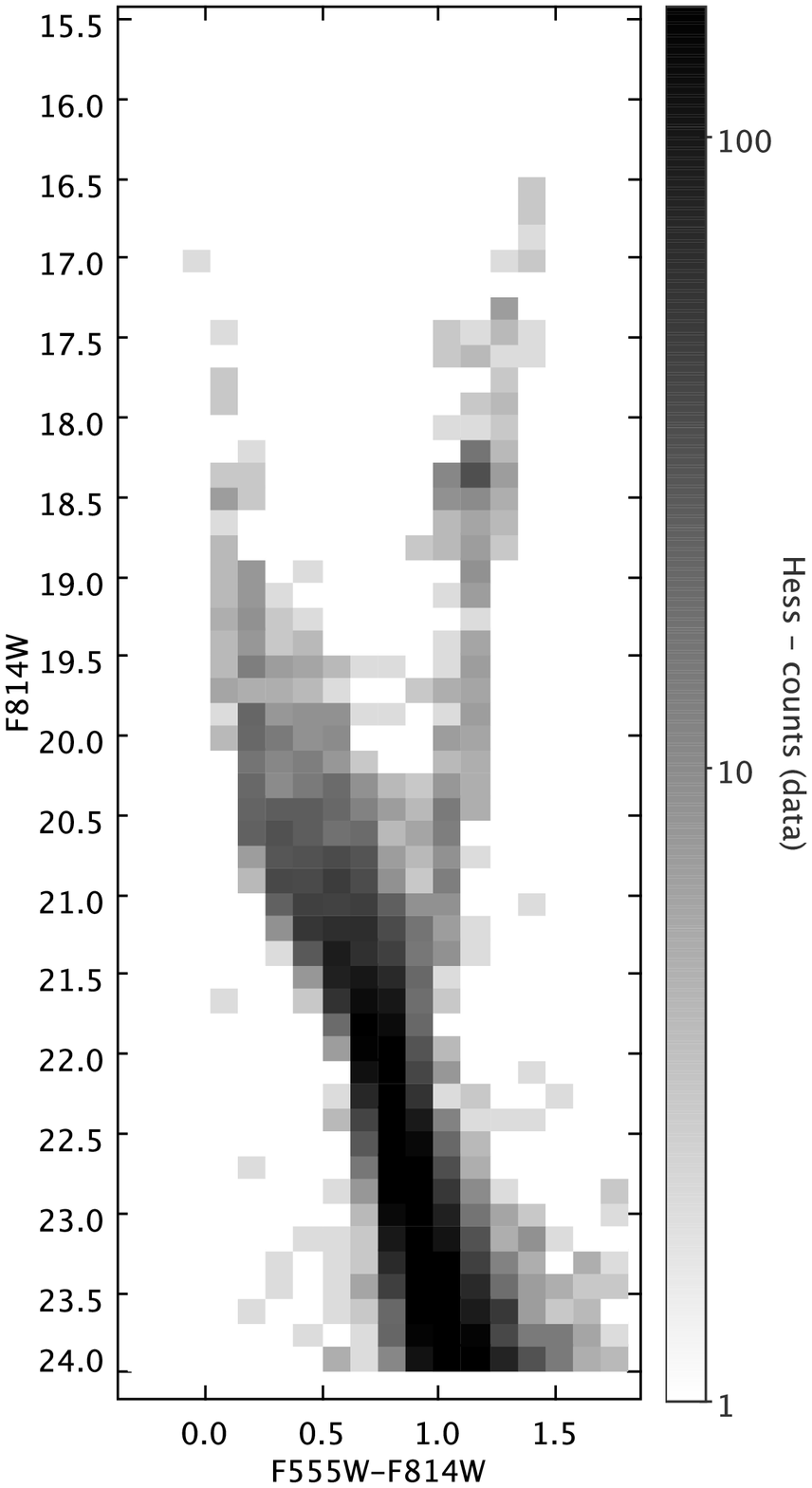}}
\resizebox{0.25\hsize}{!}{\includegraphics{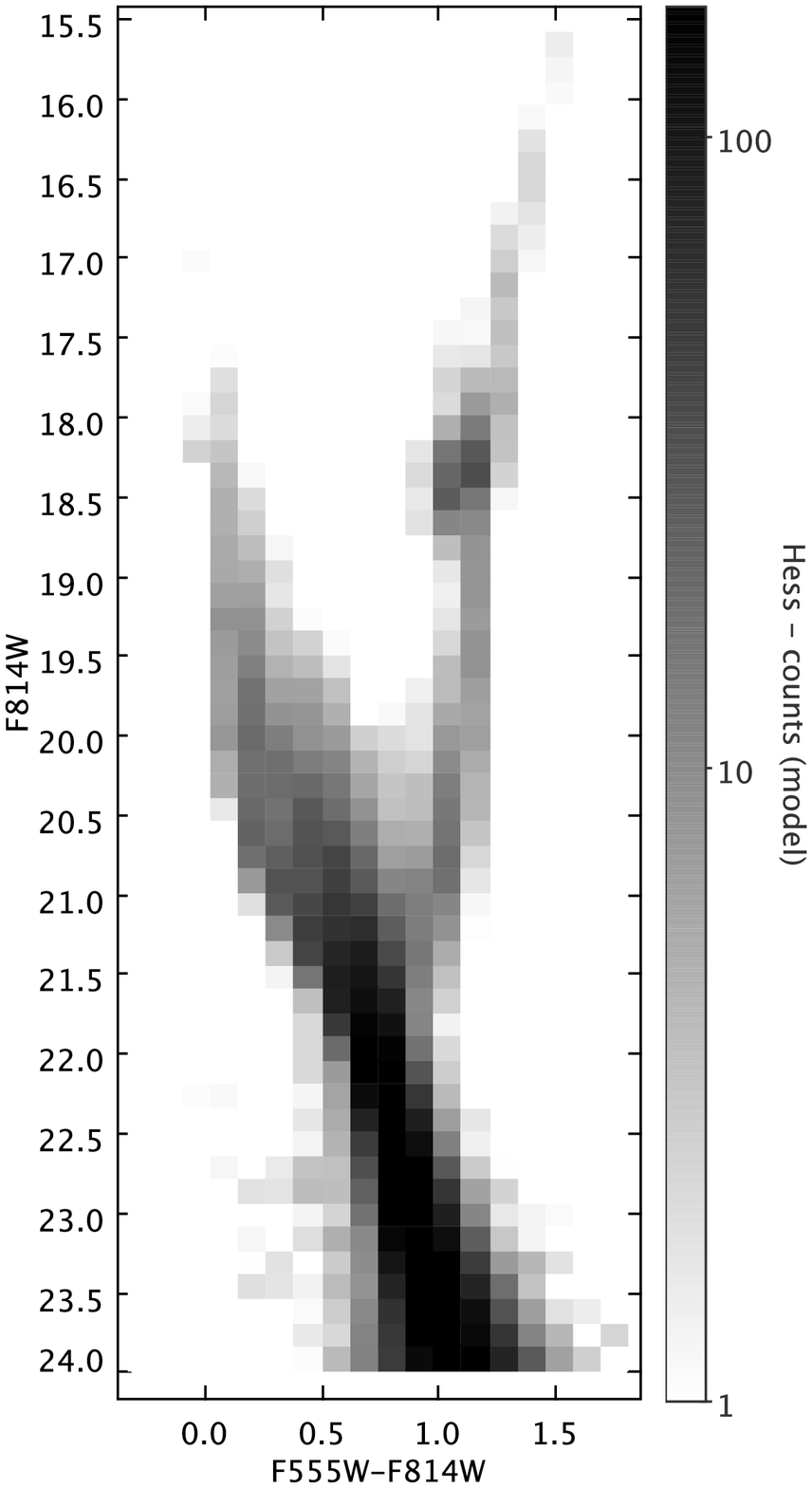}}
\resizebox{0.25\hsize}{!}{\includegraphics{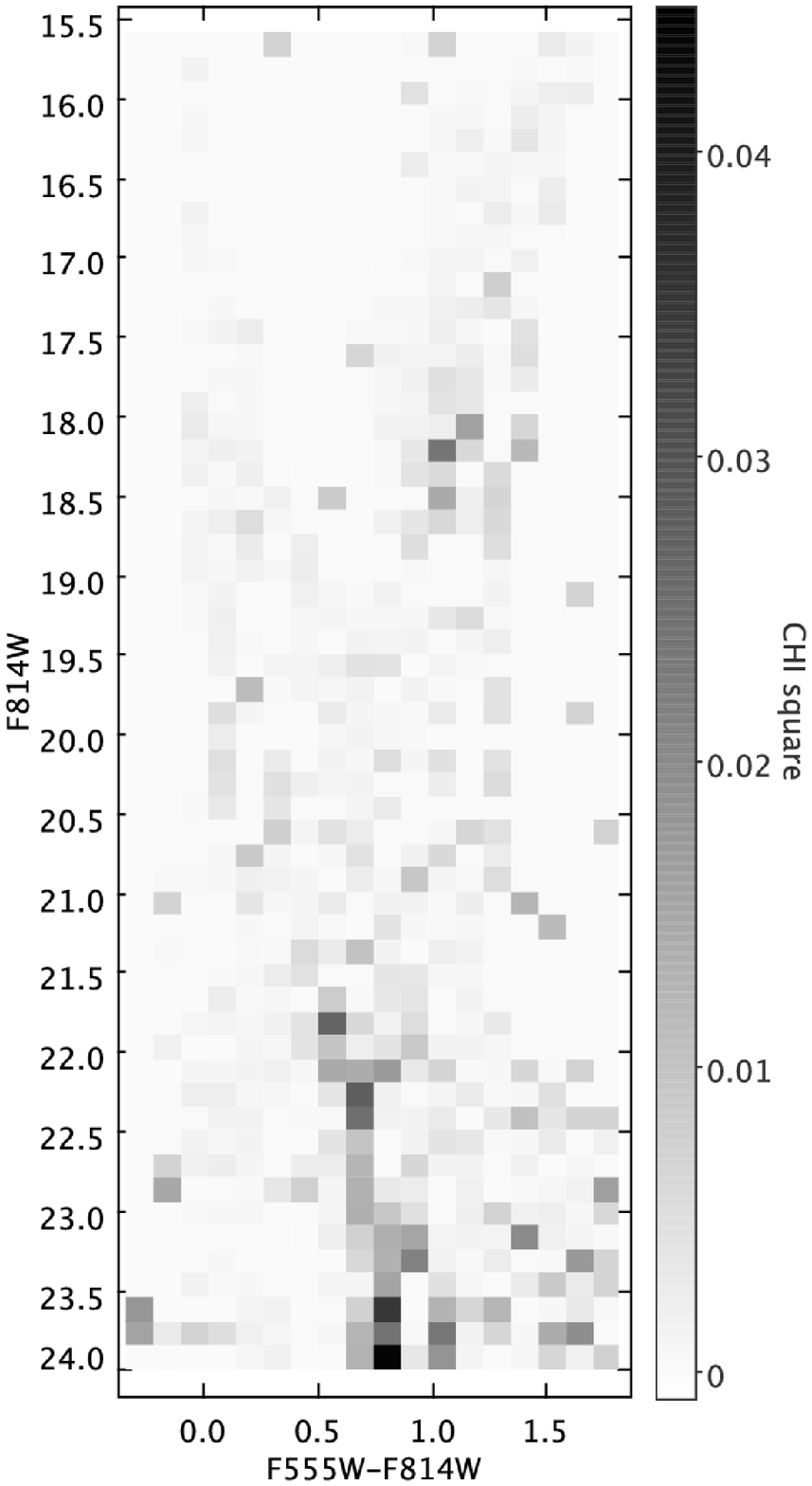}}
\caption{The Hess diagram for the NGC~1751 Field as derived from the ACS data (left panels), as recovered by the best-fitting solution (central panels), and the map of $\chi^2$ residuals (right panels). }
\label{fig_cmds_field}
\end{figure*}

\subsection{The best-fitting solution}
\label{fieldsol}

Once the database of partial models is built, we run StarFISH to find
the best-fitting solution to the observed CMDs, for a given value of
distance modulus \dmo\ and extinction \av. Both F814W$\,\times\,$
F435W$\,-\,$F814W and F814W$\,\times\,$F555W$\,-\,$F814W Hess diagrams
are used simultaneously in the process of $\chi^2$ minimization.

\begin{figure}
\resizebox{\hsize}{!}{\includegraphics{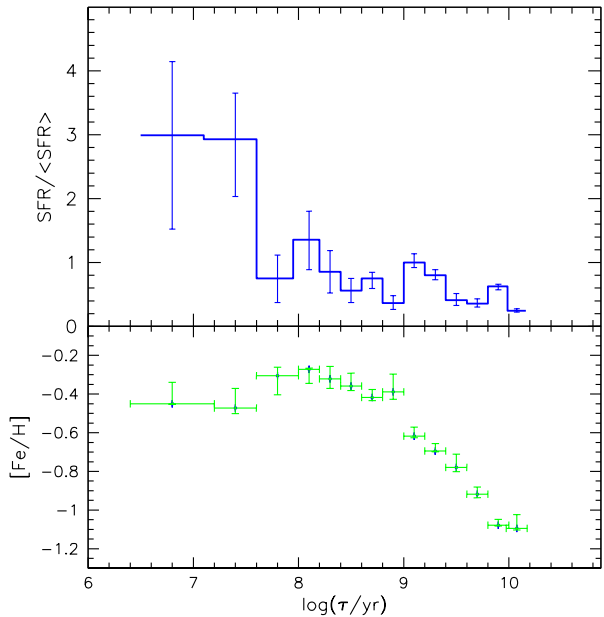}}
\caption{Top panel: best-fitting SFH for the field, together with the 
random errors ($1\sigma$). Bottom panel: the mean age--metallicity
relation.}
\label{fig_fieldSFH}
\end{figure}

\dmo\ and \av\ are then varied over the possible range of values. The
$\chi^{2}$ map of Fig.~\ref{fig_fieldchimap} shows the results in the
$\dmo\times\av$ plane. The overall best-fitting solution, with a
$\chisqmin$ of 1.6, is located at $\dmo=18.50$ and $\av=0.525$. The
$68~\%$ confidence level for this solution spans a narrow range in
distance and reddening, which is just $\Delta\dmo=0.12$~mag and
$\Delta\av=0.07$~mag wide.

Figure~\ref{fig_cmds_field} compares the Hess diagrams of the field
data (left panel) and its overall best-fitting model (right panel). It
is evident that the solutions found by StarFISH reproduce well the
observed CMD features but for the Poisson noise.

Finally, Fig.~\ref{fig_fieldSFH} presents the SFR$(t)$ and
age--metallicity relation (AMR) corresponding to this best-fitting
solution. It is remarkable that the recovered SFR$(t)$ presents
features that are consistent with those commonly found in previous
works, based on both HST data \citep{Olsen99, Holtzman_etal99,
Smecker-Hane_etal02, Javiel_etal05} and ground-based data
\citep{HZ01, HZ09}. There is an initial
burst of star formation followed by a quiescent period, with a marked
and peaked star formation for ages younger than 4~Gyr ($\log(t/{\rm
yr})=9.6$).  In particular the peaks of star formation at
approximately 1.5~Gyr ($\log(t/{\rm yr})=9.2$), 500~Myr ($\log(t/{\rm
yr})=8.7$), 100~Myr ($\log(t/{\rm yr})=8.0$) and 10~Myr ($\log(t/{\rm
yr})=7.0$) are in tight agreement with those found by
\citet{HZ09} for the global SFH of the LMC.  Concerning the AMR, the
result for the NGC~1751 field is consistent with those derived from
the LMC stellar clusters \citep{Kerber_etal07, HZ09} and for the LMC
field \citep{Carrera_etal08}.

\section{The SFH for NGC~1751}
\label{sec_clusterSFH}

\subsection{Overview of NGC~1751 parameters from literature}

As for the LMC field, also for the NGC~1751 cluster we need a set of
physical parameters to start with the SFH-recovery work. They are
based on the following works:

The cluster metallicity as determined by the Ca~{\sc ii} method is of
$\feh=-0.44\pm0.05$ \citep{Grocholski_etal06}, which is a typical
value for an intermediate-age LMC cluster
($\feh=-0.48\pm0.09$; \citealt{Grocholski_etal06}).

\citet{Milone_etal08} identified a double MSTO in the HST/ACS F435W 
vs.\ F435W$\,-\,$F814W CMD for this cluster. Using isochrone fitting,
they determined ages between 1.3 and 1.5 Gyr, a distance modulus of
18.45~mag, $\ebv=0.22$ ($\av\simeq0.70$), and a metallicity of
$Z=0.008$ ($\feh\simeq-0.38$). 

\citet{Milone_etal08} also determine a binary fraction $f_{\rm b}$ of 
$0.13\pm0.1$ for mass ratios $q$ larger than 0.6 in NGC~1751. Despite
the small error bar quoted by them, their estimate is admitedly a
preliminary one. The careful determination from \citet{Elson_etal1998}
for the LMC cluster NGC~1818, finds $f_{\rm b}$ values varying from
$\sim 0.20$ to $\sim 0.35$ as one goes from the cluster center to the
outer parts. We adopt here the conservative value of $f_{\rm b}=0.2$
for $q>0.7$. Our previous results for NGC~419 \citep{Rubele_etal10}
indicate that the results of the SFH recovery do not depend
significantly on the choice of binary fraction.

As for the extinction, the reddening maps from the Magellanic Clouds
Photometric Survey \citep[MCPS; ][]{Zaritsky_etal04} and
\citet{PejchaStanek09} provide discrepant values for the
NGC~1751 direction. Within 3~arcmin from the cluster, MCPS gives
$\av=0.47\pm0.34$ for hot stars, and $\av=0.59\pm0.39$ for cool
stars. From the same dataset, \citet{Pessev_etal08} determined
$\av=0.65 \pm 0.06$. \citet{PejchaStanek09} instead find
$\left<E_{V\:-\:I}\right>=0.150\pm0.293$ ($\left<\av\right> =
0.293\pm0.062$), although their value is based on just 5 stars within
a $2^\circ\times2^\circ$ area.

The distance modulus to the LMC disk in the NGC~1751 direction is of
about 18.55~mag, as revealed by independent methods: \citet[][AGB
stars]{vdMC01}; \citet[][red clump stars]{OlsenSalyk02};
\citet[][Cepheid stars]{Nikolaev_etal04}; \citet[][RC stars from
MCPS]{SubramanianSubramaniam10}.

The above-mentioned works provide the initial guesses for the many
cluster parameters to be determined in this work.

\subsection{The partial models for NGC~1751}
\label{sec_partialmodels}

For NGC~1751 we assume a constant age-metallicity relation, i.e., a
single value for the metallicity for all ages, since so far there are
no evidences for significant spreads in metallicity in such star
clusters \citep[e.g.][]{Mucciarelli_etal08, Rubele_etal10}. The age
interval covered by our partial models goes from $\log(t/{\rm
yr})=8.9$ to $9.4$, which is much wider than the interval suggested by
the position of NGC~1751 MMSTOs. So, for each set of parameters, we
have a total of 10 partial models, completely encompassing the age
interval of interest.  We have explored 5 metallicity values, going
from $\feh$ from $-0.75$ to $-0.35$ at steps of 0.1~dex. For each one
of these mean $\feh$ values, a small metallicity spread of 0.02~dex is
assumed.
       
This definition of partial models might already be good enough to our
aims to find the best-fitting solution for the cluster
centre. However, we know that every portion of the ACS/WFC image is
contaminated from the LMC field, and that this field contamination is
well evident in the observed CMDs (especially for the
Ring). Therefore, it is quite tempting to add, to the above-mentioned
list of partial models, an additional one corresponding to the LMC
field. The hope is that this partial model will allow StarFISH to
properly fit the field contamination across the CMDs on NGC~1751,
hence improving the fitting of cluster itself. 

The inclusion of a partial model for the field is a novelty of this
work, and is suggested as an alternative to the commonly used method
of field star decontamination \citep[see e.g.,][]{Milone_etal08,
BonattoBica07}, which consists in subtracting from the cluster CMD the
stars with colors and magnitudes similar to the ones in the field {\em
before} deriving the cluster parameters. The advantage of our method
is that the field becomes an integral part of the $\chi^2$ and error
analysis; the latter is performed using the correct number statistics
-- i.e. considering the Poisson noise from the field, which is
certainly present in cluster data -- without implying any change in
the method already set for these tasks.

There are then two different ways at which this partial model for the
field can be built.  (1) The simplest one is that of taking the
observed Hess diagrams for the field region (left panels in
Fig.~\ref{fig_cmds_field}). This diagram however is affected by the
Poisson fluctuations in the numbers of stars, so that it might not
describe in a suitable way the field actually observed in other parts
of the ACS images. (2) The second alternative is to use the Hess
diagram built from the best-fitting solution of the field (right
panels in Fig.~\ref{fig_cmds_field}) which is obviously much more
continuous and smooth over the CMDs than the observed one. This model
has another clear advantage: the Hess diagram can be easily re-built
using the output SFH for the field {\em together} with the ASTs
derived for the cluster Centre or Ring. In this way, we are able to
simulate the field under the same conditions of crowding met in the
cluster data. We indeed adopt this latter alternative in the
following.

\begin{figure*}
\begin{minipage}{0.3\textwidth}
\resizebox{\hsize}{!}{\includegraphics{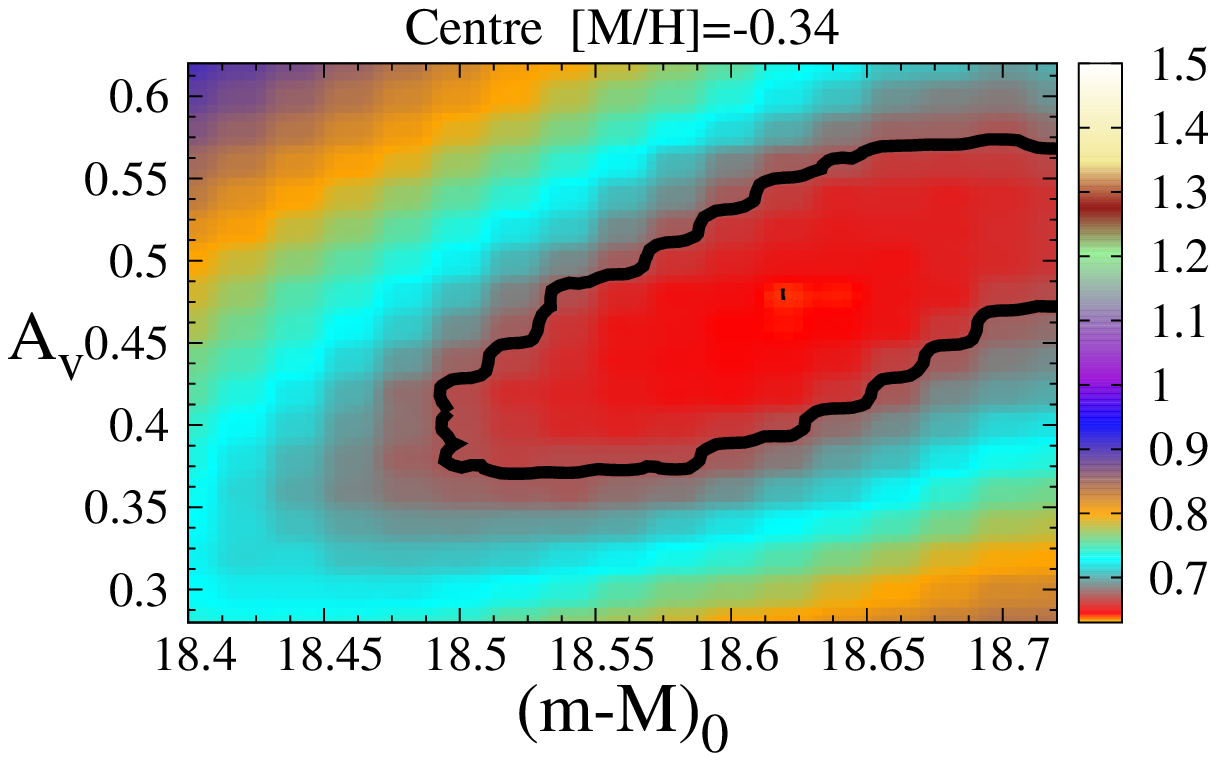}}
\resizebox{\hsize}{!}{\includegraphics{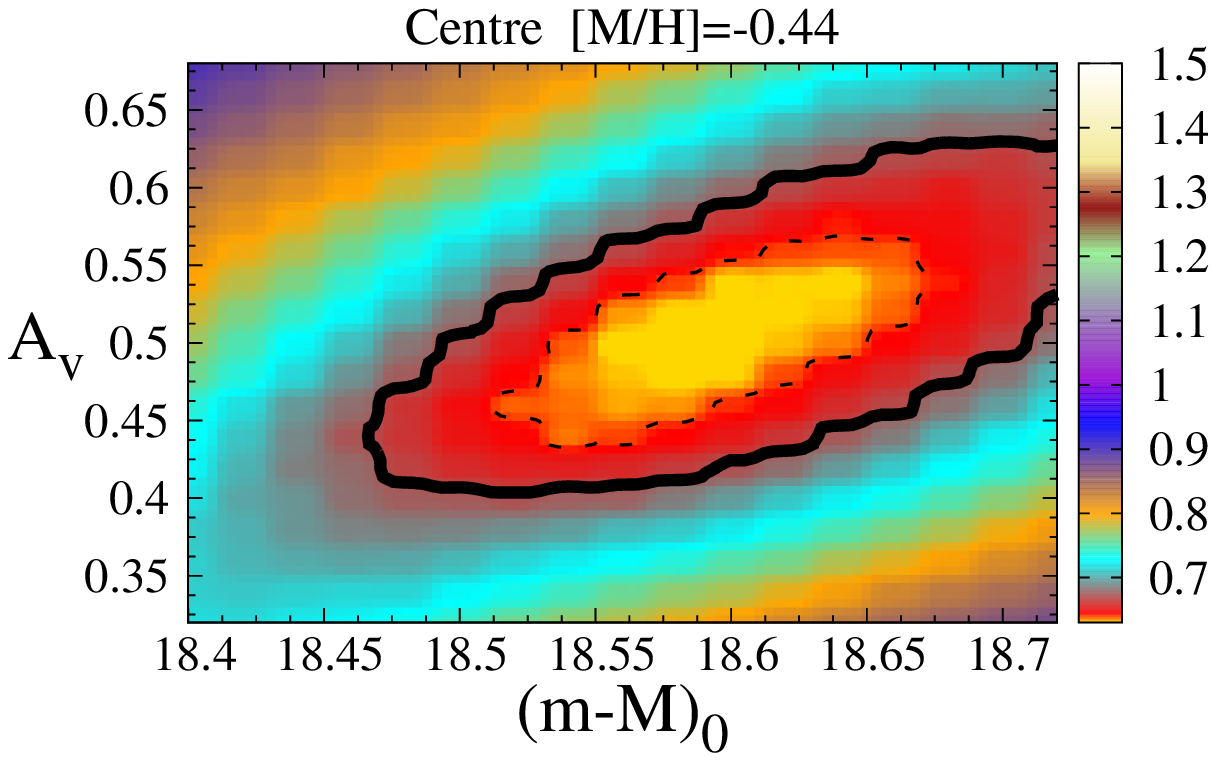}}
\resizebox{\hsize}{!}{\includegraphics{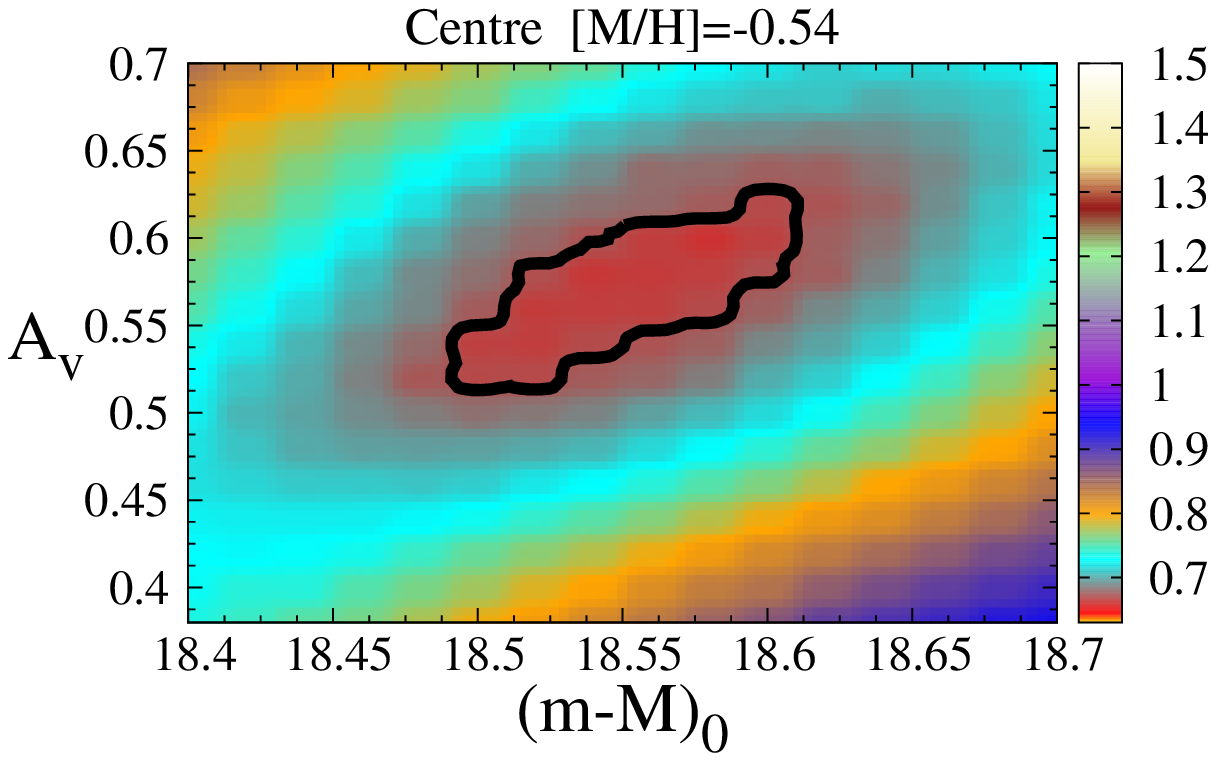}}
\resizebox{\hsize}{!}{\includegraphics{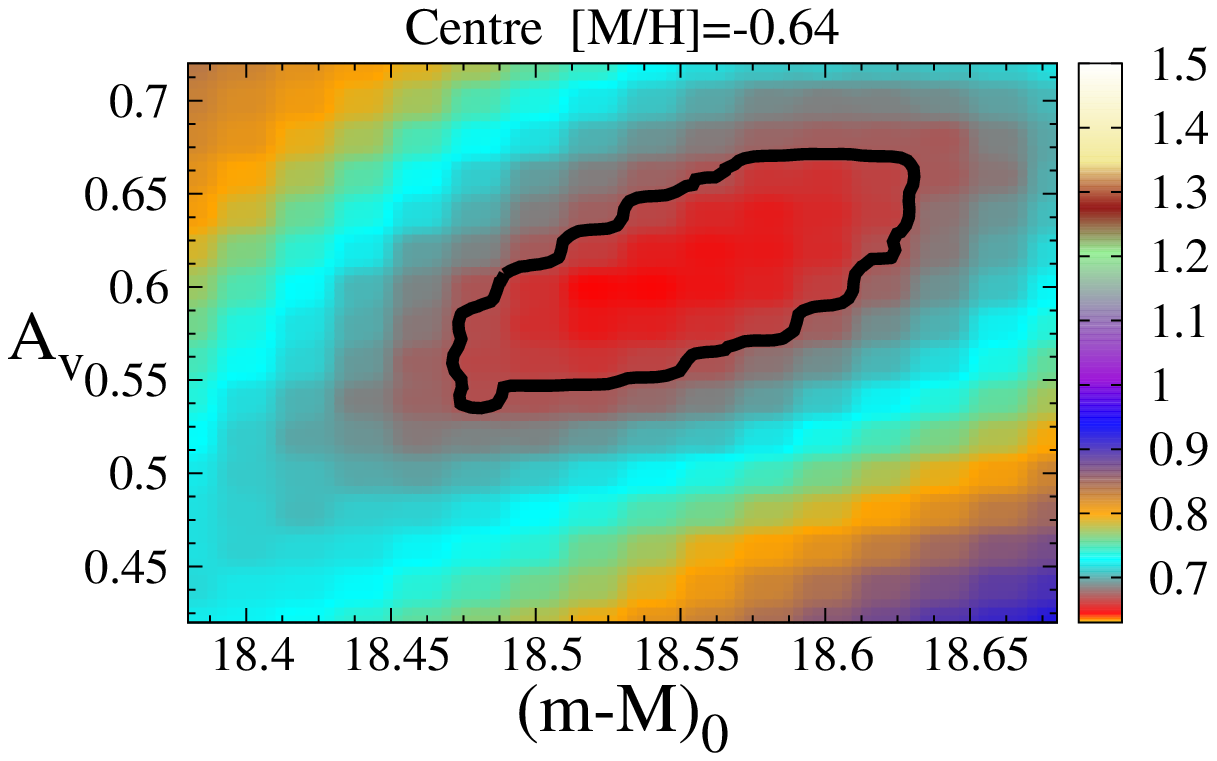}}
\resizebox{\hsize}{!}{\includegraphics{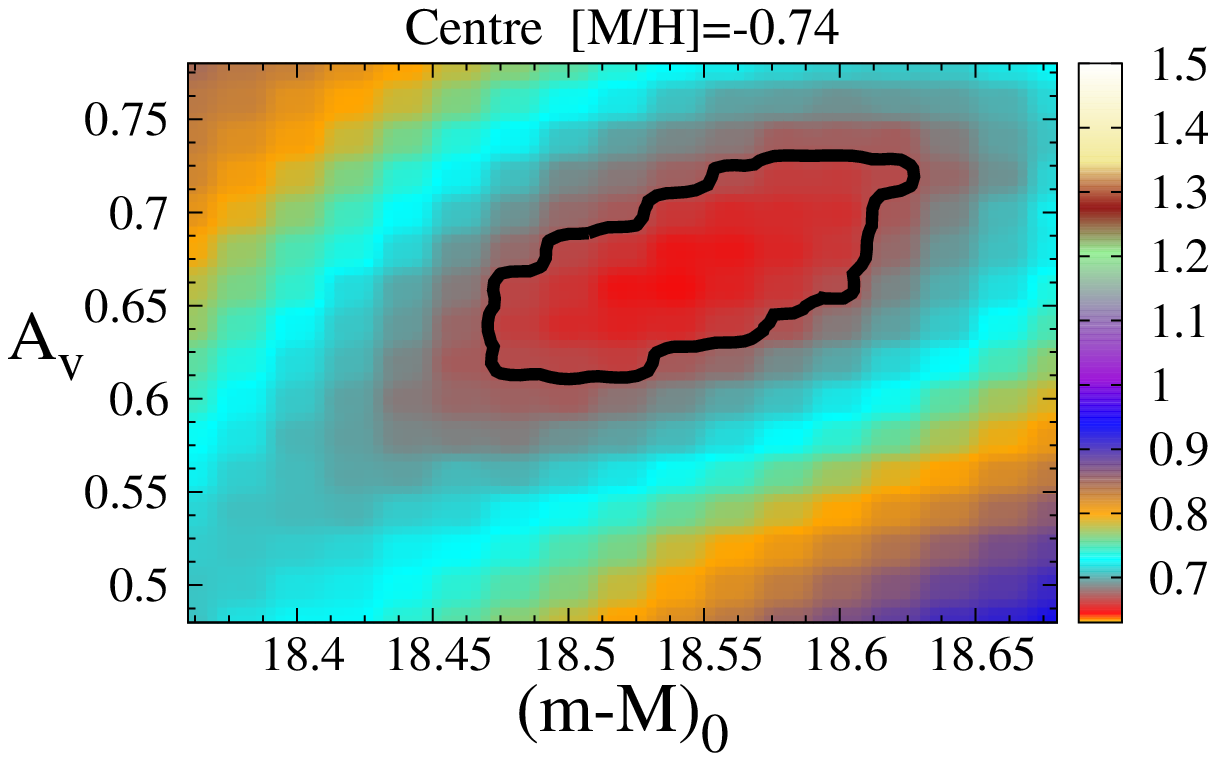}}
\end{minipage}
\begin{minipage}{0.3\textwidth}
\resizebox{\hsize}{!}{\includegraphics{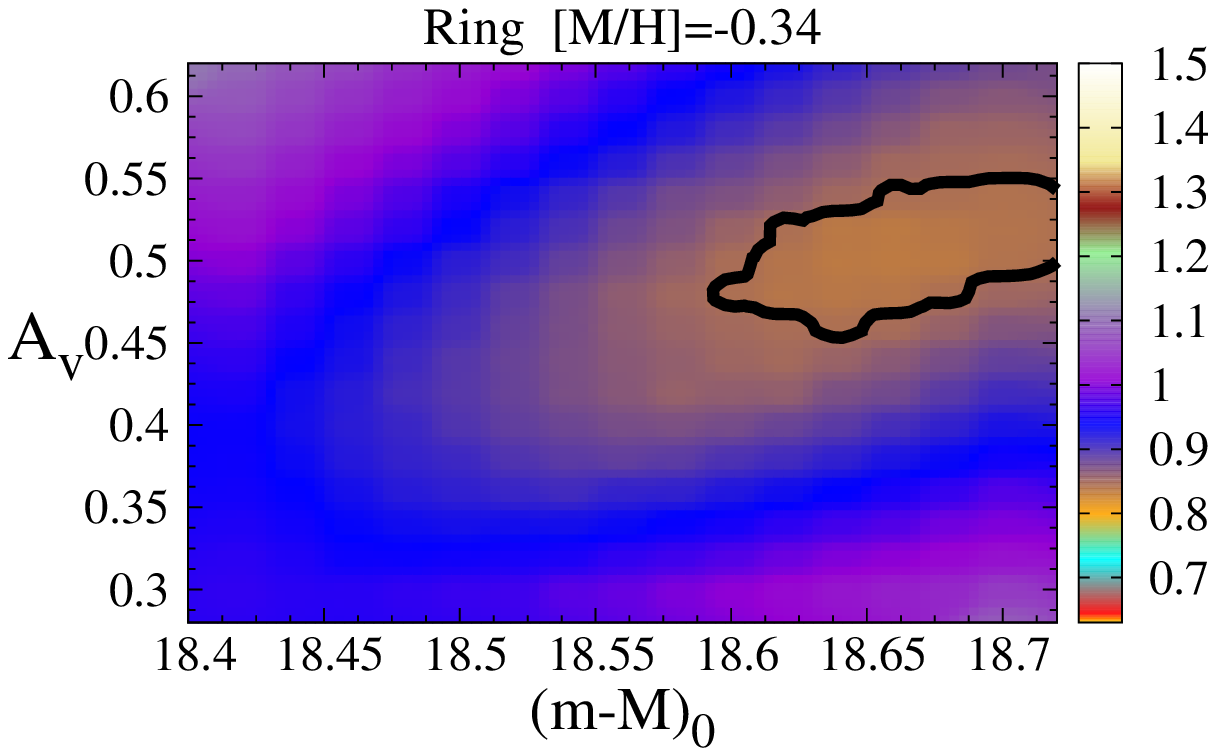}}
\resizebox{\hsize}{!}{\includegraphics{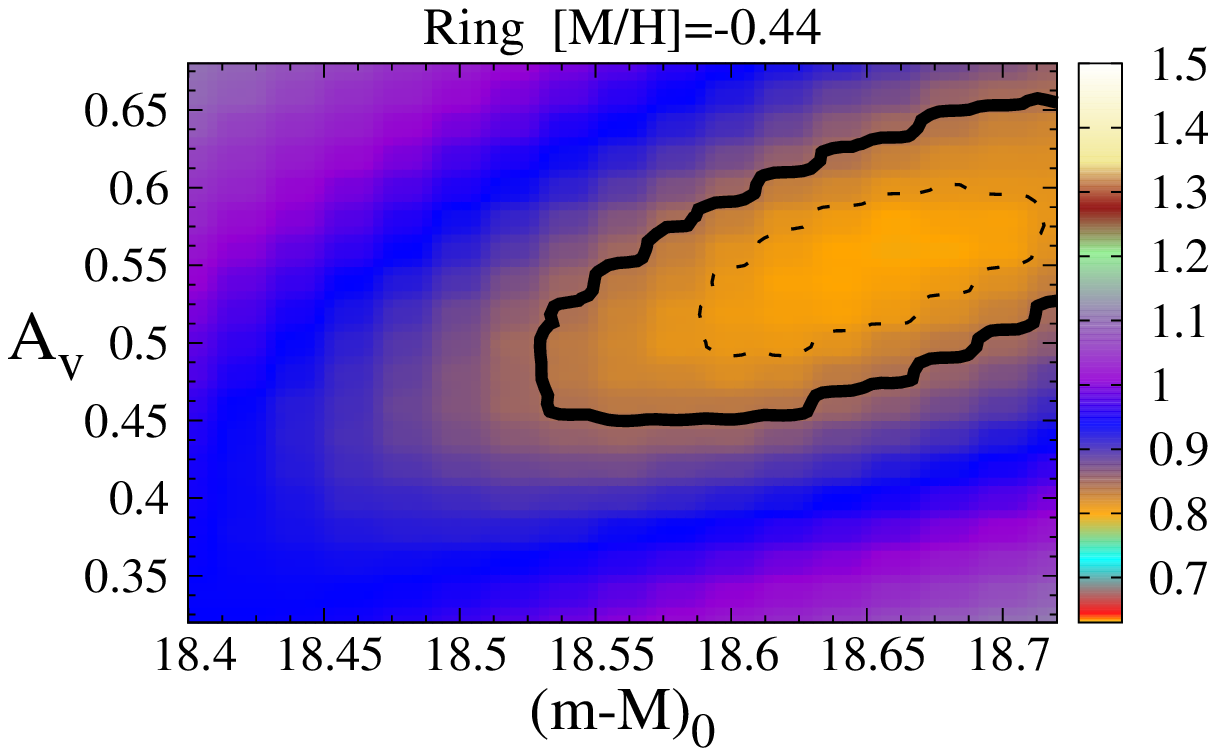}}
\resizebox{\hsize}{!}{\includegraphics{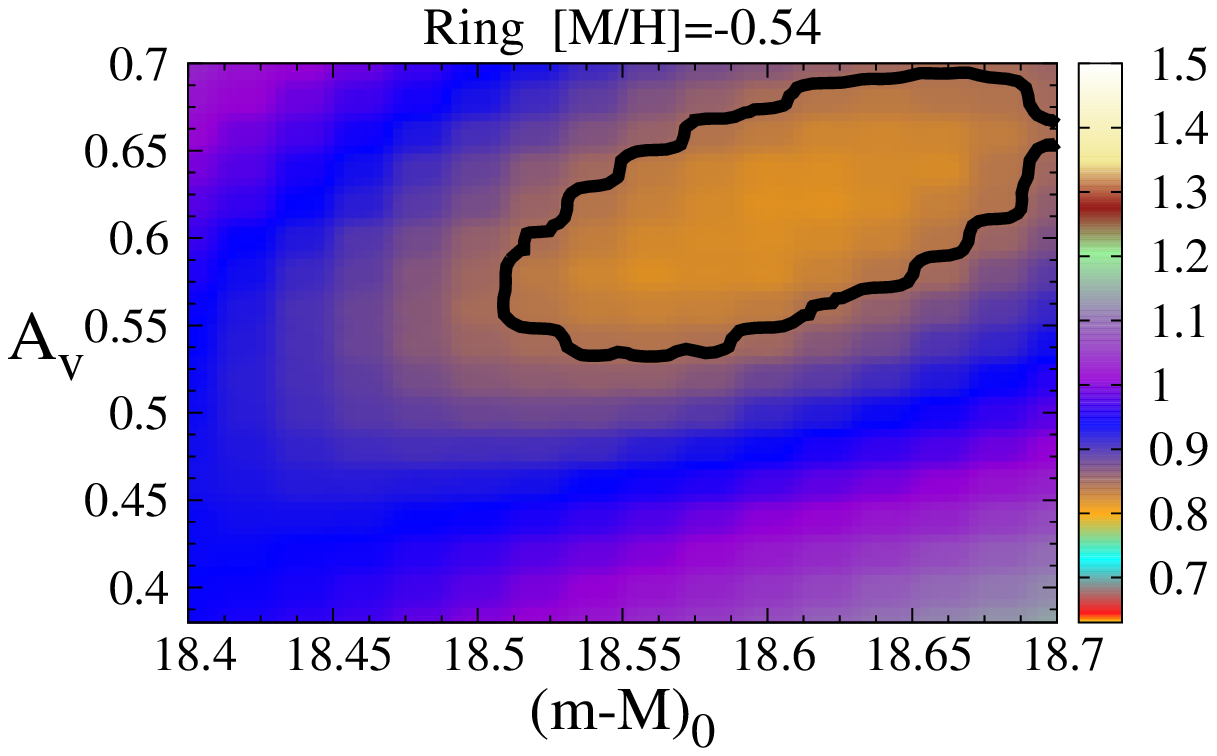}}
\resizebox{\hsize}{!}{\includegraphics{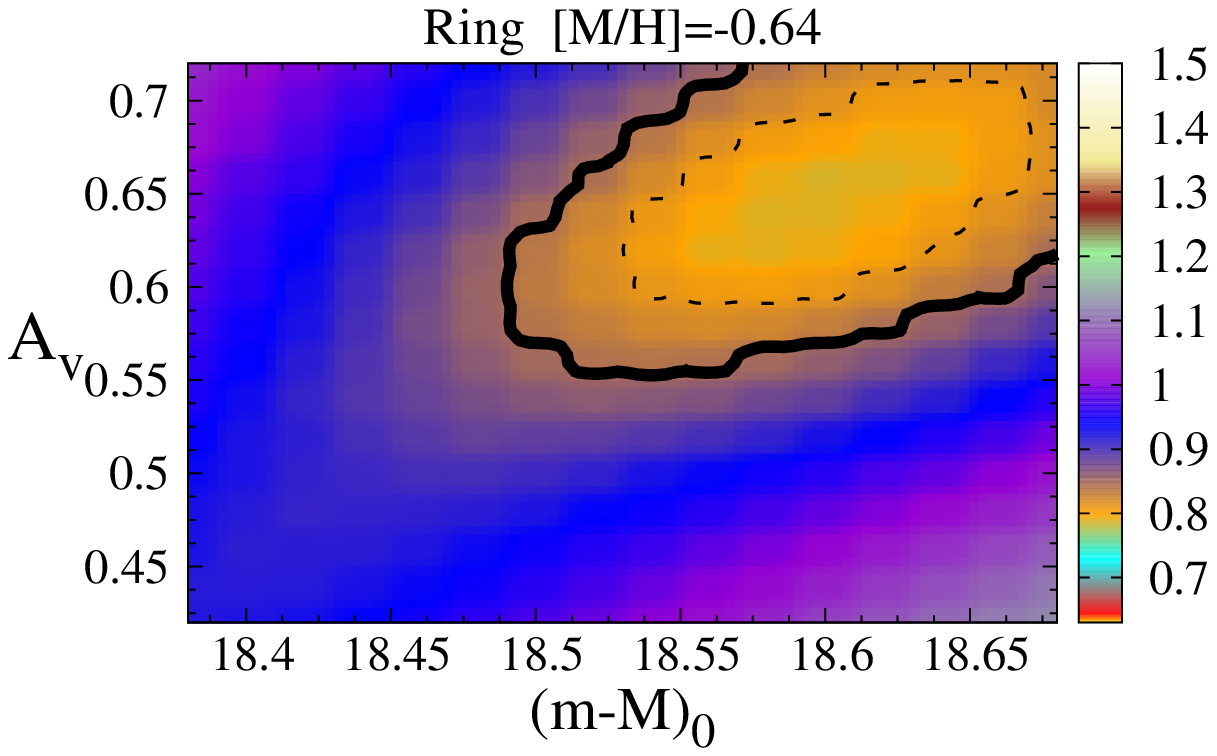}}
\resizebox{\hsize}{!}{\includegraphics{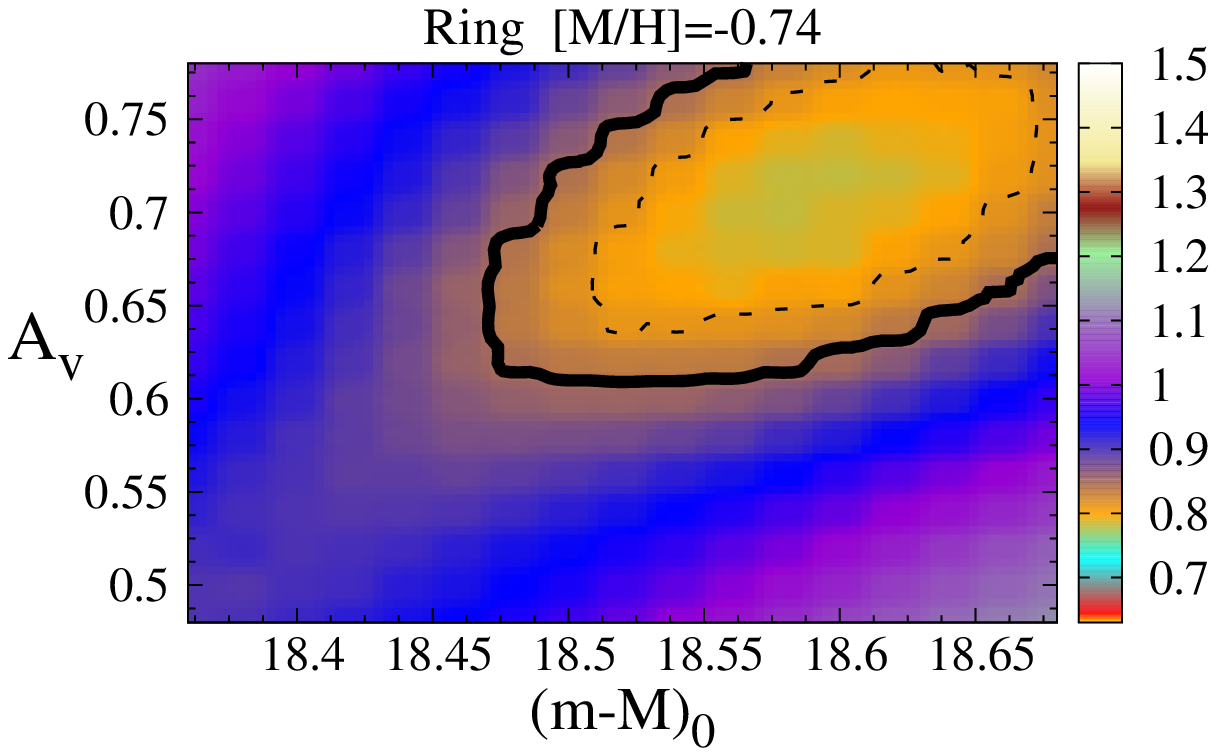}}
\end{minipage}
\caption{Maps of the $\chisqmin$ obtained from the SFH-recovery, as a 
function of $\dmo$ and $\av$, for several $\feh$ values (from $-0.34$
to $-0.74$ at steps of $-0.1$~dex, from top to bottom) and for both
the cluster Centre and Ring (left and righ panels, respectively). The
black lines delimit the regions within a 68~\% (continuous line) and
95~\% confidence levels (dotted lines) of the absolute best solution,
which is found at $-0.34$~dex for the Centre, and at $-0.74$~dex for
the Ring. The $\chisqmin$ for the Centre best solution is of 0.625.}
\label{chi2_map}
\end{figure*}

\begin{figure*}
\resizebox{0.33\hsize}{!}{\includegraphics{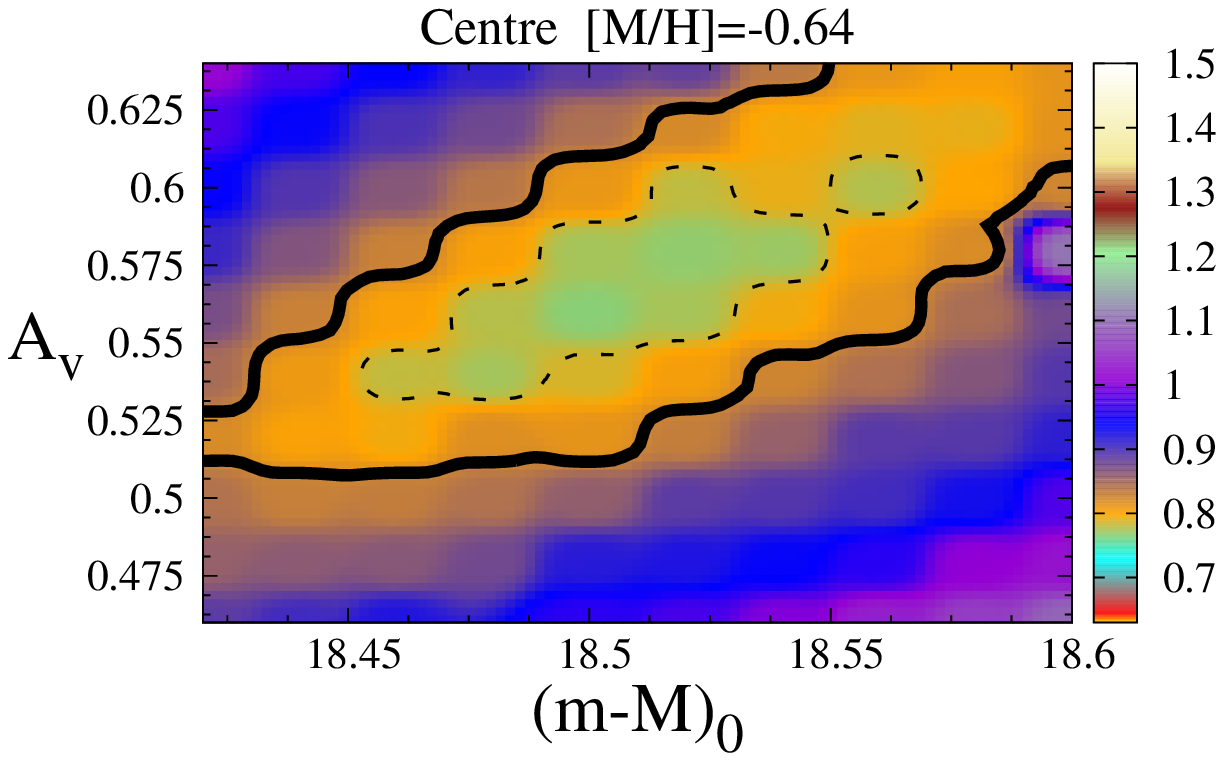}}
\resizebox{0.33\hsize}{!}{\includegraphics{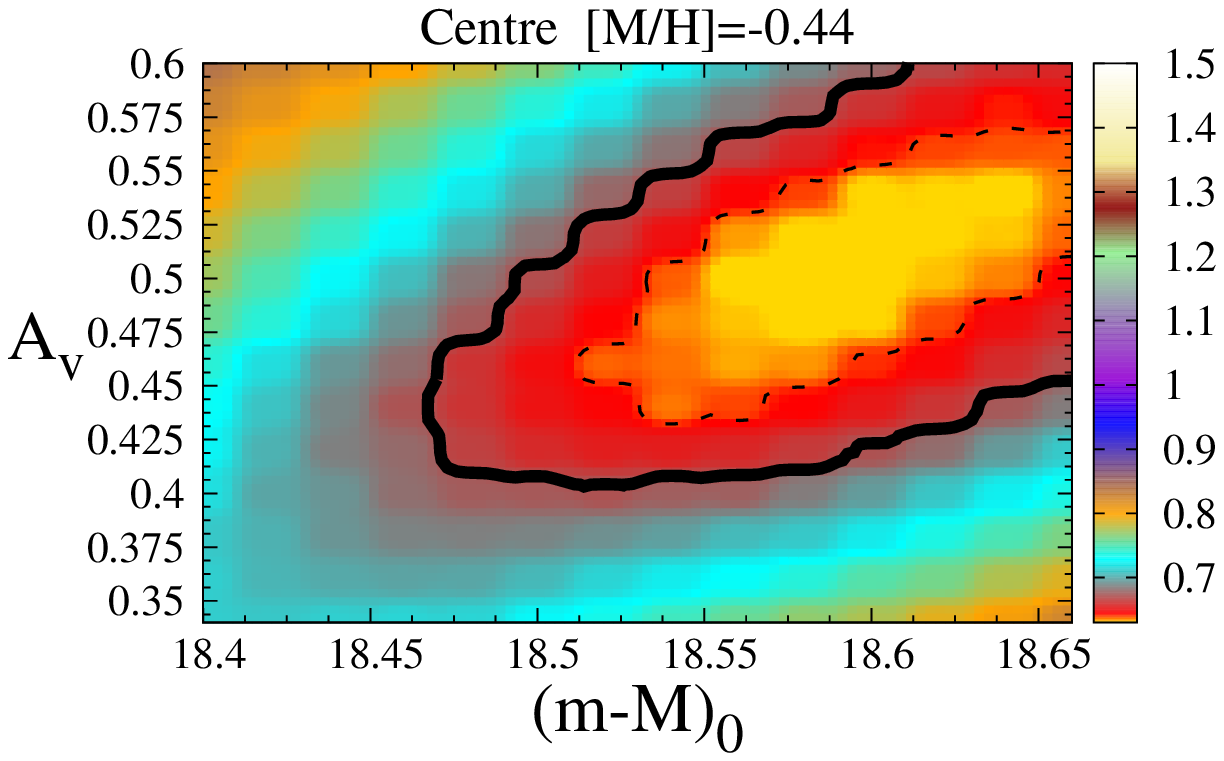}}
\resizebox{0.33\hsize}{!}{\includegraphics{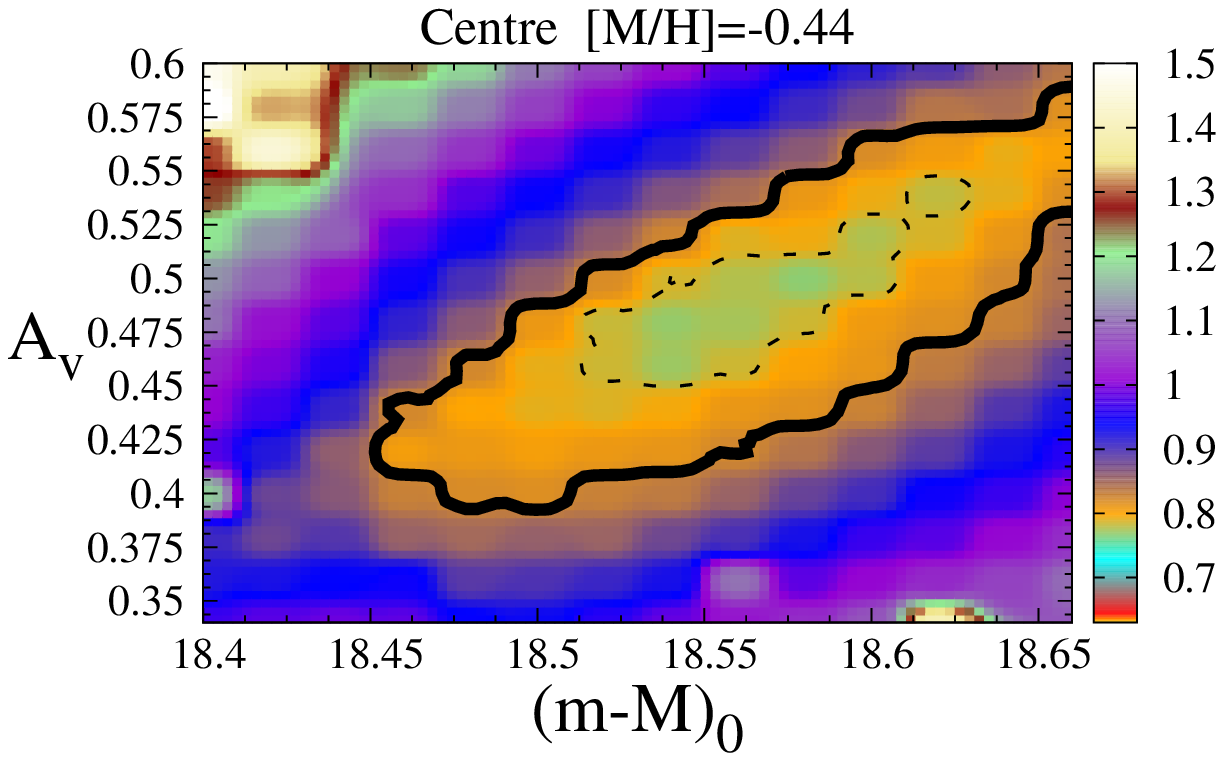}}
\caption{Maps of the $\chisqmin$ obtained during SFH-recovery in the 
Centre region, as a function of \dmo\ and \av. The {\bf left panel}
shows the map for the best-fitting metallicity of $\feh=-0.64$,
obtained in the case A (i.e. not taking into account a partial model
for the LMC field). The minimum $\chisqmin$ is of 0.77. The {\bf
middle panel} shows the same for case~B (i.e. using the LMC field
partial model) and $\feh=-0.44$, which is the best-fitting metallicity
in this case. It is evident that these solutions are characterized by
a significantly smaller level of $\chisqmin$ overall, with a minimum at
$0.62$. For comparison, the {\bf right panel} shows the best-fitting
solutions for case A and $\feh=-0.44$. Also in this case, the
$\chisqmin$ are significantly higher (and very close to the one in the
letfmost panel). }
\label{chi2_mapwftmp}
\end{figure*}

\subsection{The SFH for the cluster Centre}

\begin{figure*}
\begin{minipage}{0.74\hsize}
\resizebox{0.33\hsize}{!}{\includegraphics{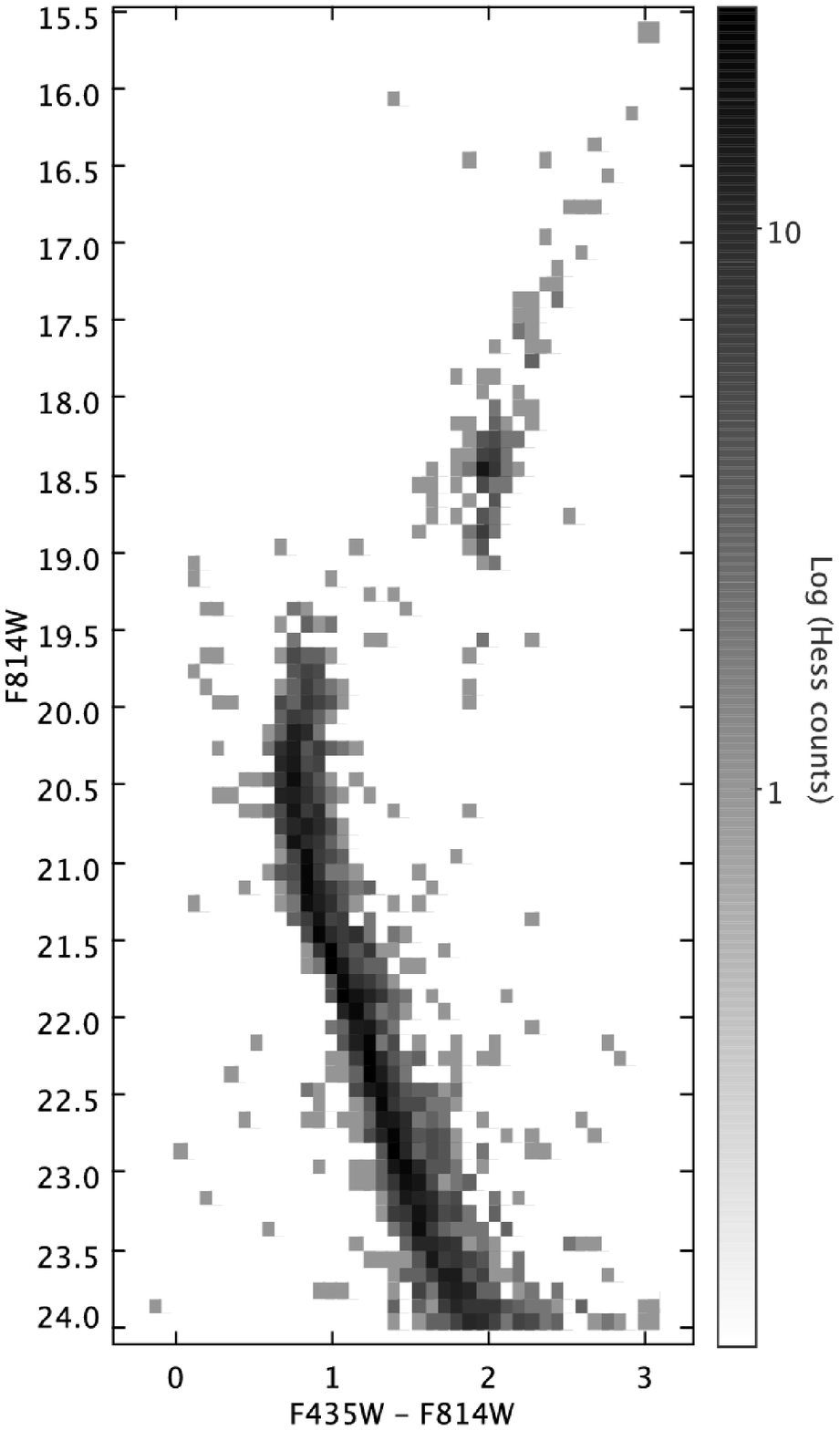}}
\resizebox{0.33\hsize}{!}{\includegraphics{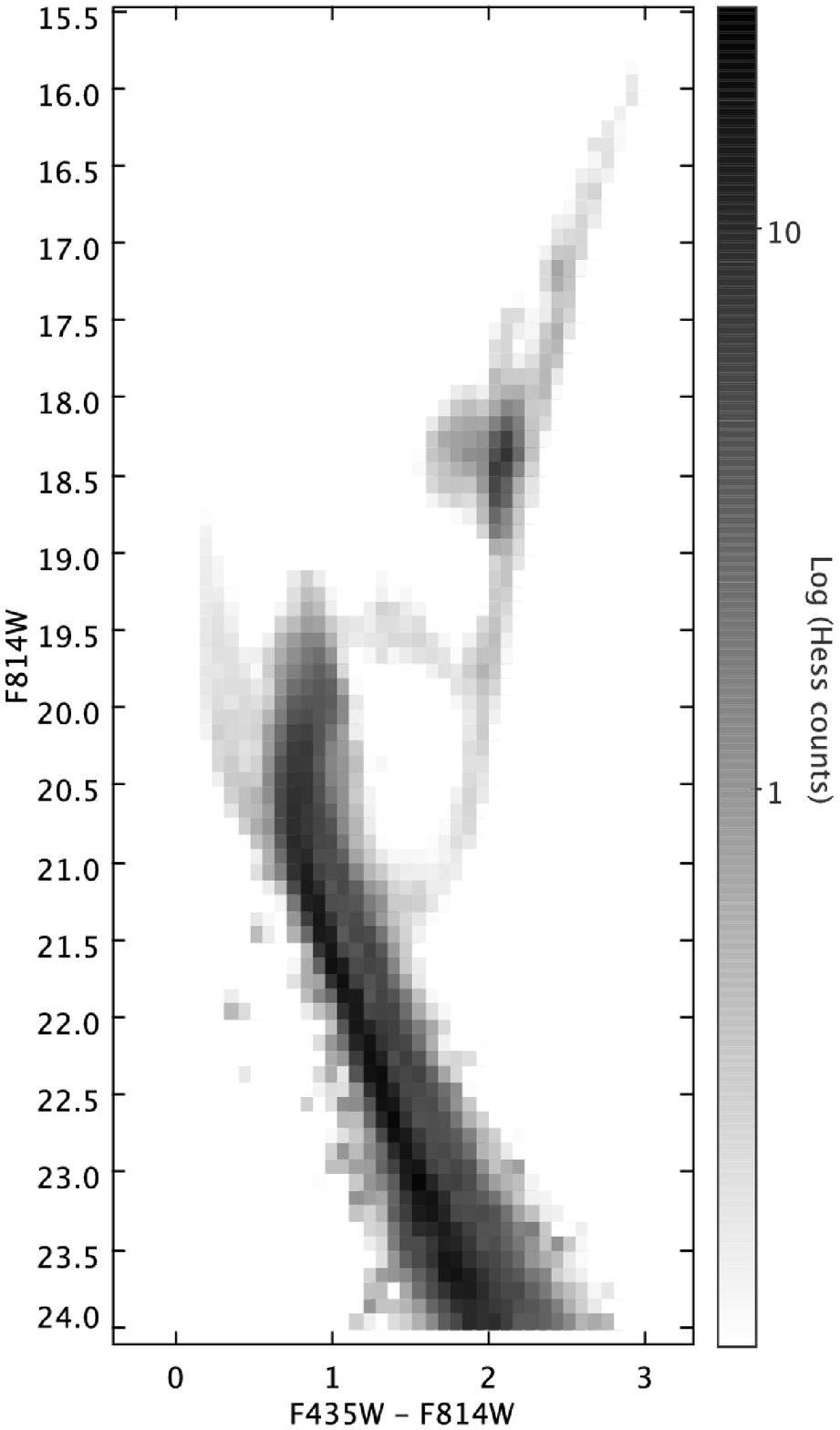}}
\resizebox{0.33\hsize}{!}{\includegraphics{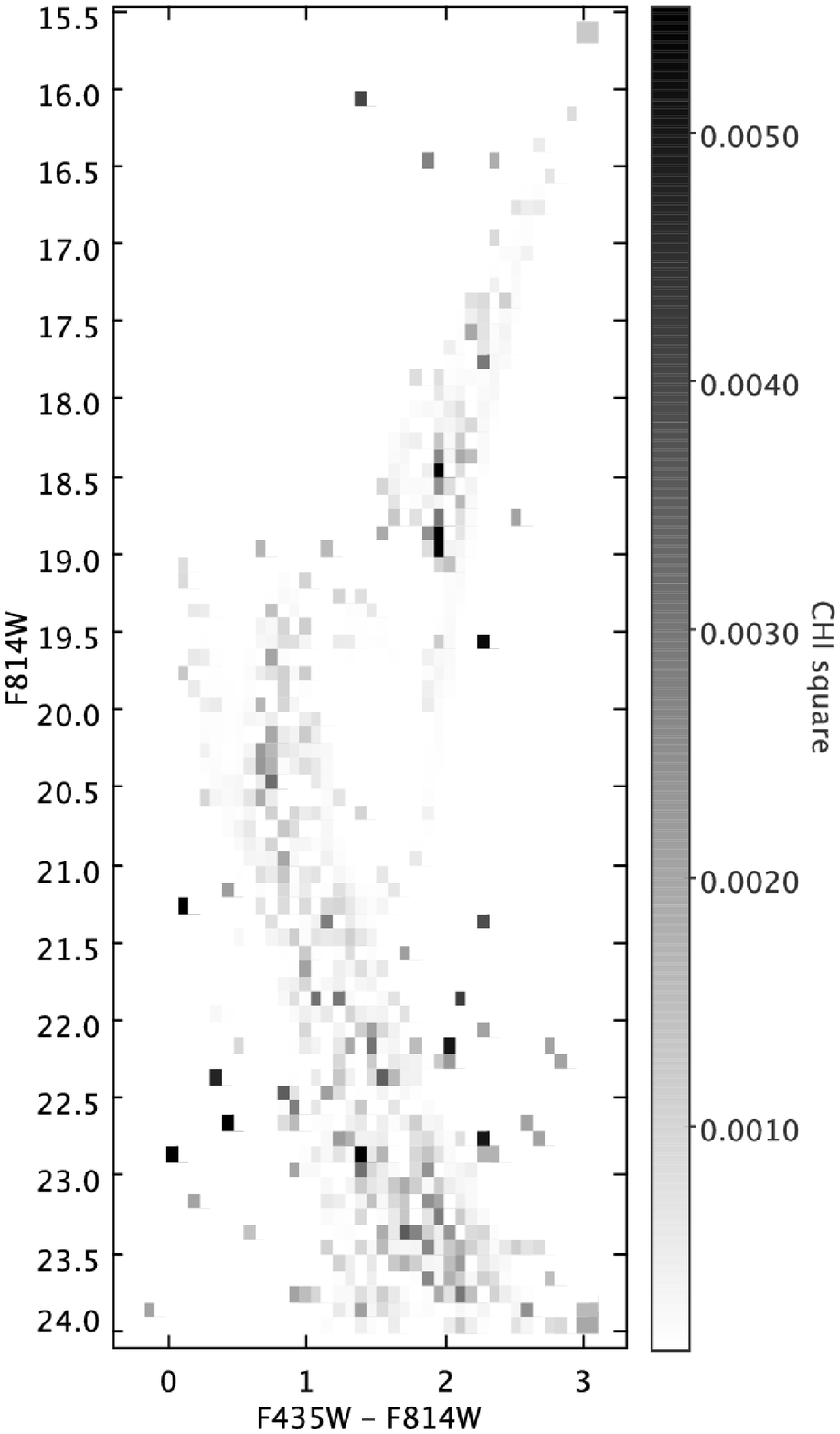}}

\resizebox{0.33\hsize}{!}{\includegraphics{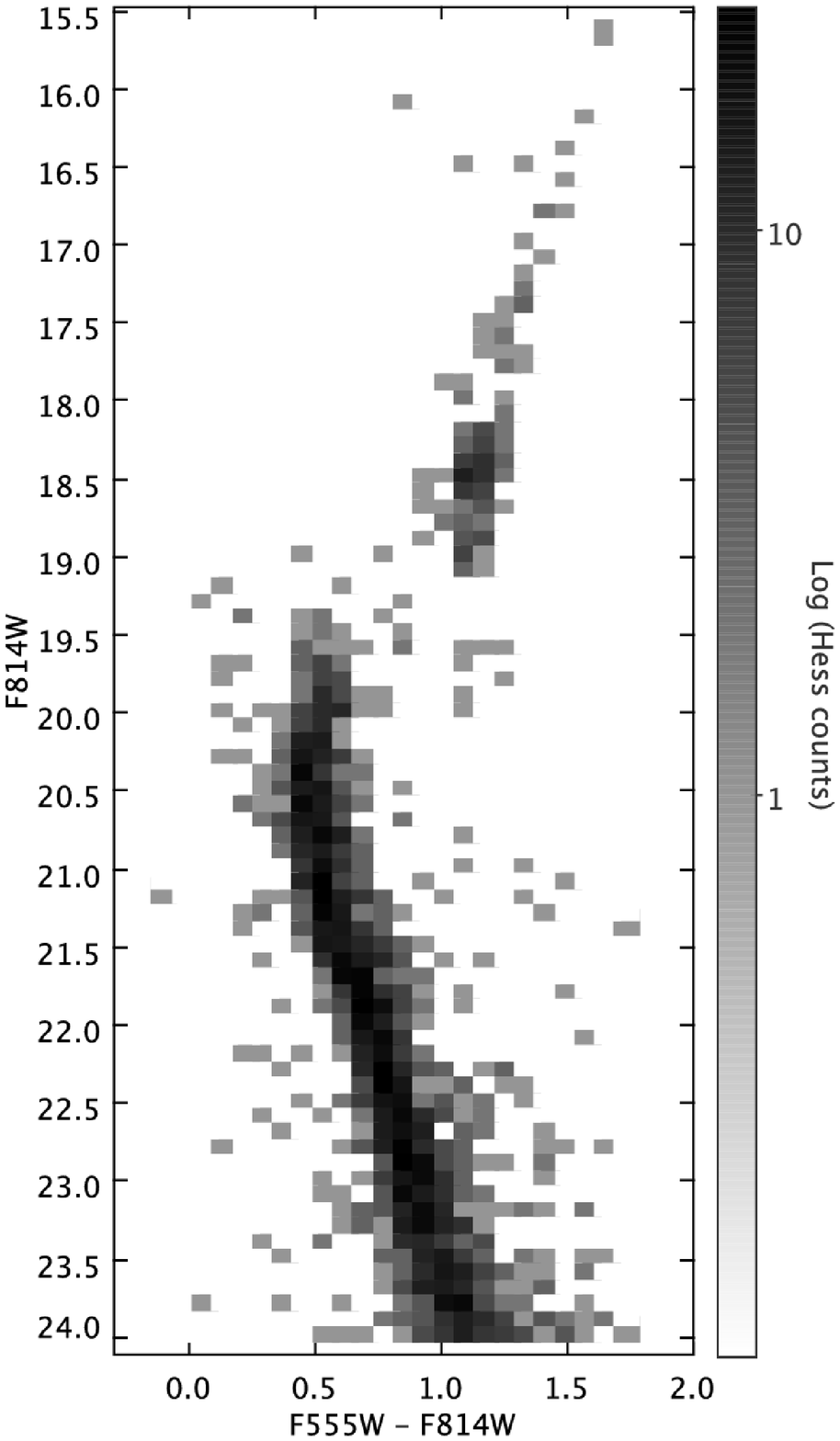}}
\resizebox{0.33\hsize}{!}{\includegraphics{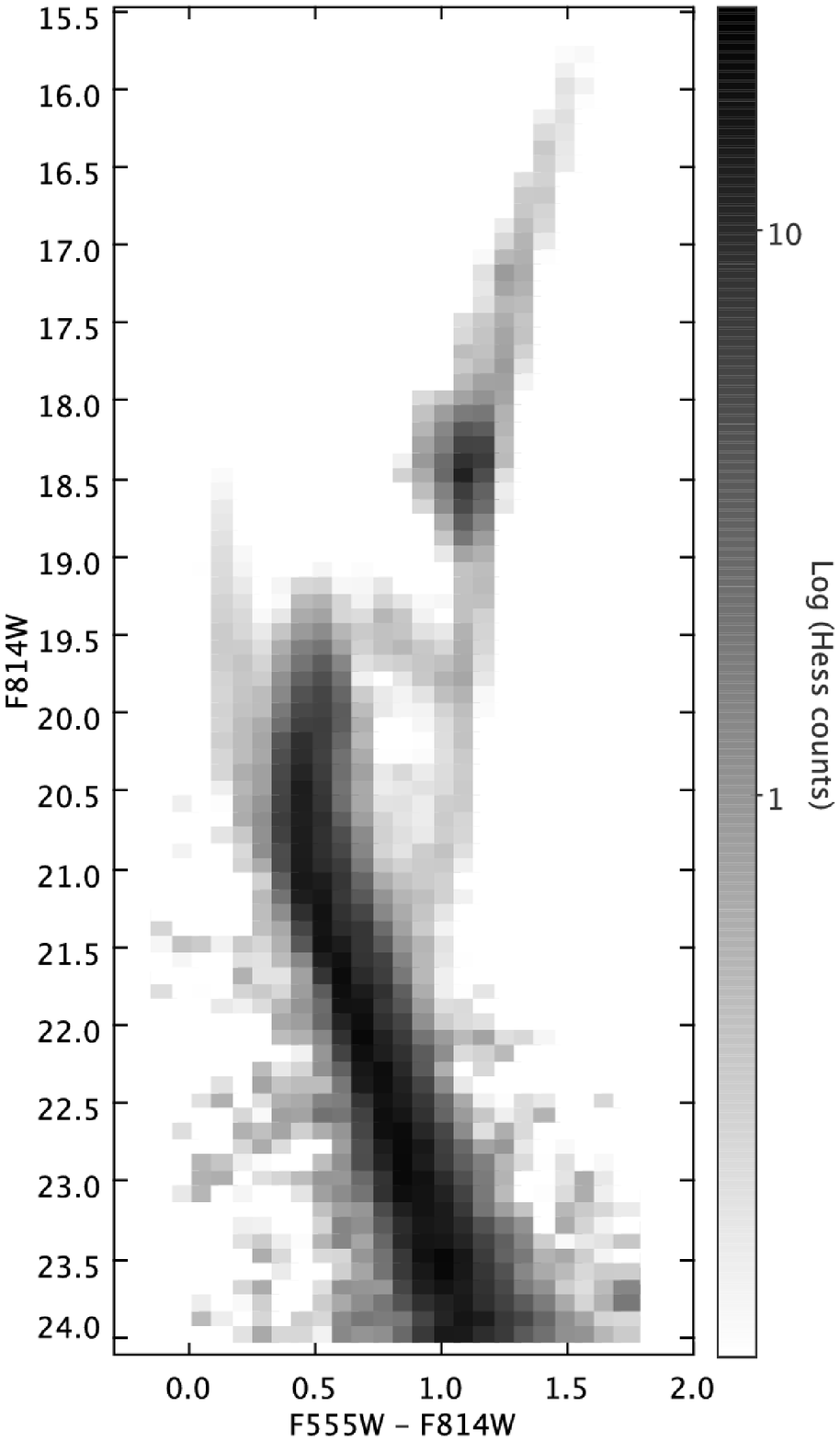}}
\resizebox{0.33\hsize}{!}{\includegraphics{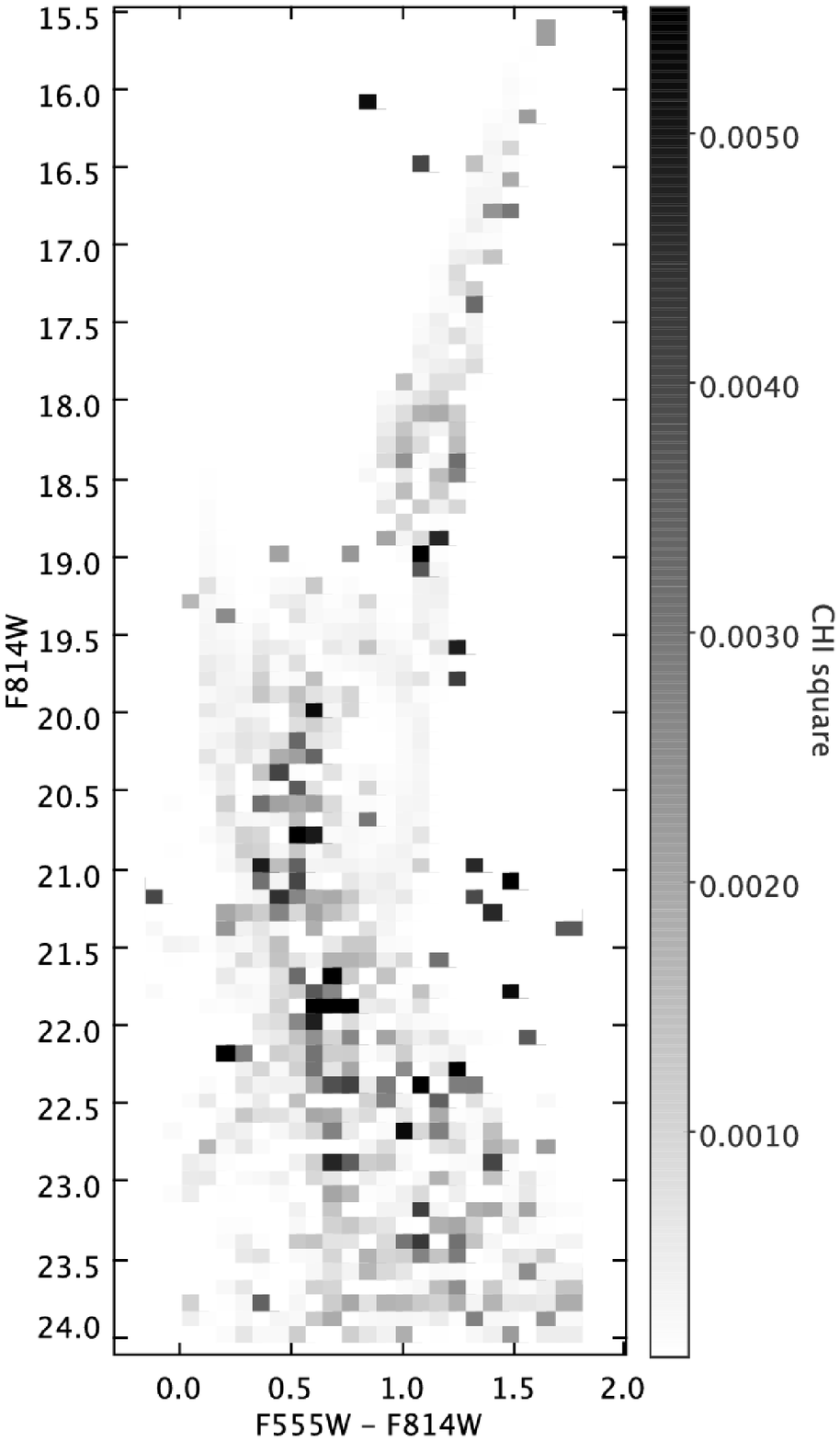}}
\end{minipage}
\caption{The Hess diagrams for the NGC~1751 Centre data (left panels), its 
best-fitting solution model (central panels), and the $\chi^2$ map
(right panels). The top panels are for the F435W$\,-\,$F814W vs. F814W
diagrams, the bottom ones for F555W$\,-\,$F814W vs. F814W.}
\label{data_model}
\label{fig_residuals}
\end{figure*}

\begin{figure*}
\resizebox{0.49\hsize}{!}{\includegraphics{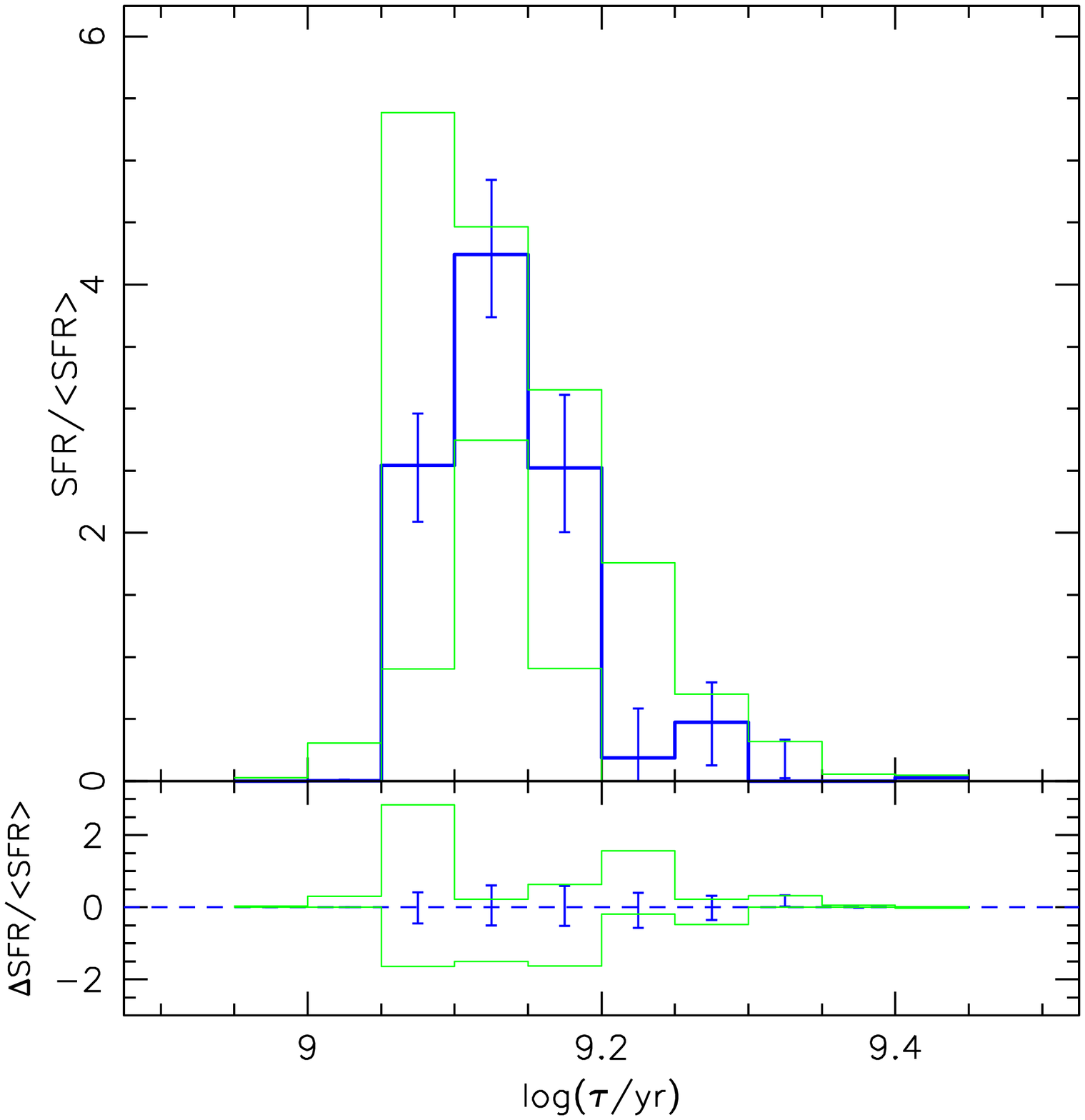}}
\resizebox{0.49\hsize}{!}{\includegraphics{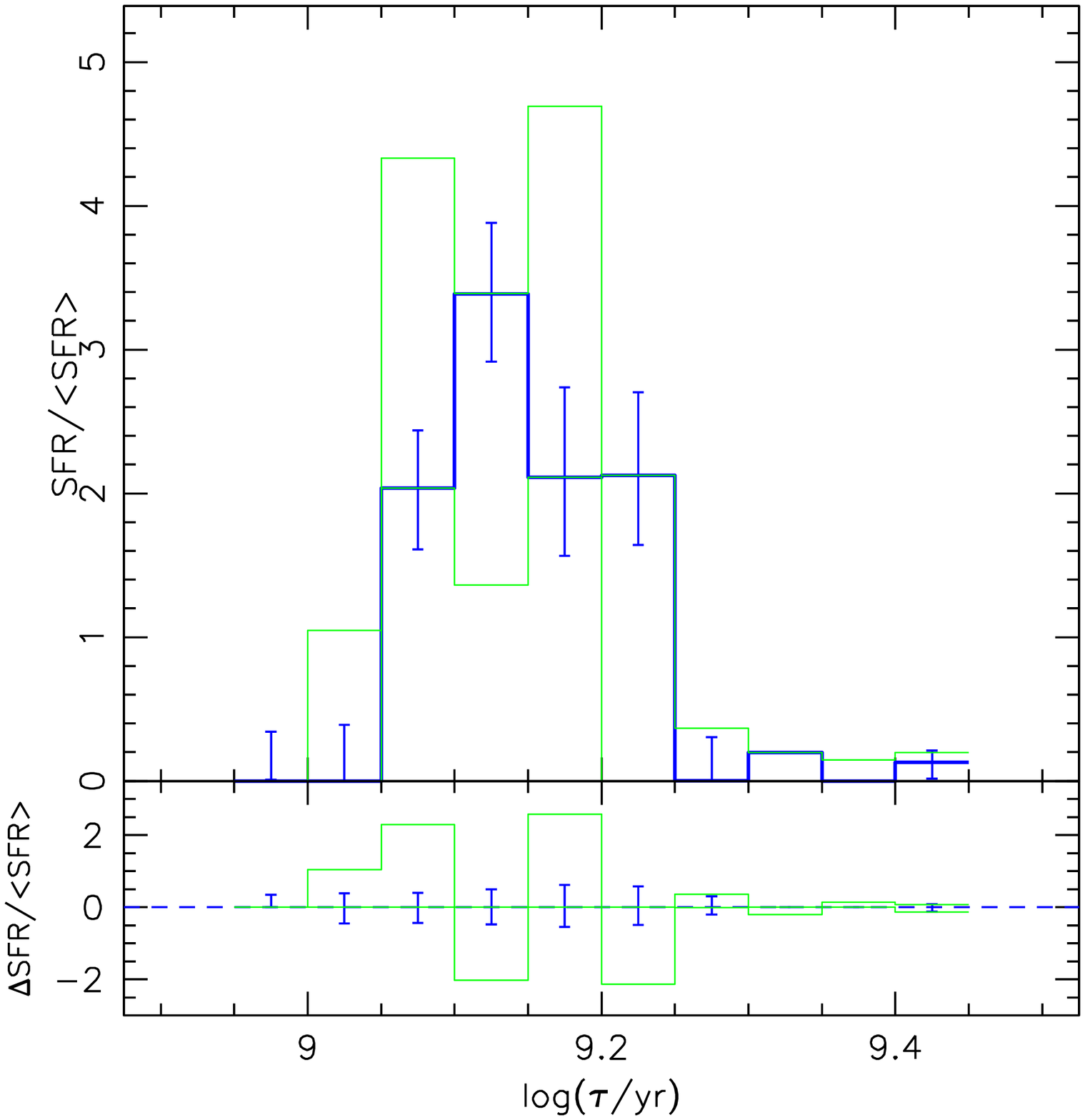}}
\caption{The blue lines show the SFR$(t)$ for the cluster Centre (left 
panel) and for the Ring (right panel). The error bars are
random errors. The green lines indicate the systematic errors, inside
the area of 68~\% confidence level in Fig.~\ref{chi2_map}.}
\label{sfr}
\end{figure*}
 
\subsubsection{Results with and without the LMC field}

The SFH-recovery is performed assuming a given set of \dmo, \av, and
\feh\ values and fixing the binary fraction at a value of 0.2 in the
case of cluster models.  In order to limit the space of parameters to
be covered, the procedure is essentially the following: for a given
value of \feh, we perform SFH-recovery for each point in a grid
covering a significant region of the \dmo\ versus \av\ plane, so as to
build a map of the $\chisqmin$ for the solutions. Examples of these
maps are presented in Fig.~\ref{chi2_map}. The maps are extended
enough so that the minimum $\chisqmin$ for a given value of \feh\ can
be clearly identified, as well as the regions in which $\chisqmin$
increases by a factor of about 1.5. The typical resolution of such
maps is of 0.02~mag in \dmo\ and 0.02~mag in \av.

Let us first start discussing the case of the cluster
Centre. Fig.~\ref{chi2_mapwftmp} shows the maps of $\chisqmin$ as a
function of \dmo\ and \av, for two series of SFH-recovery experiments
made under very similar conditions, i.e. using the same data and
ASTs. The only difference is that in some cases (hereafter {\bf
case~A}), we do not use the partial model for the LMC field in
StarFISH, whereas in other cases we do it (hereafter {\bf
case~B}). The result of considering the LMC field is quite evident:
although in both A and B cases the best-fitting solution is found for
about the same value of \dmo\ and \av, in case~B the $\chisqmin$
values are systematically smaller, which means better overall fits of
the CMDs. Moreover, it is evident that in case~A the presence of the
LMC field falses the determination of the best-fitting cluster
metallicity: indeed, in case~A the best-fitting model of $\feh=-0.64$
is found to be slightly favoured over the one with $\feh=-0.44$. In
case~B, instead, the best-fitting solution at $\feh=-0.44$ is clearly
favoured. Notice that, at the $\sim1.5$~Gyr old ages of NGC~1751, the
field is found to have a metallicity of about $-0.65$ (see
Fig.~\ref{fig_fieldSFH}), which probably helps, in case~A, to move the
$\chisqmin$ minimum to $\feh=-0.64$.

These experiments demonstrate that even a small fraction of field
contamination may affect significantly the results of SFH-recovery, if
not properly taken into account. In the following, we adopt case~B as
the default, since it demonstrately takes the LMC field into account
and improves the quality of the final results for the Centre of
NGC~1751.

\subsubsection{Characteristics of the best-fitting solution}

Complete maps of $\chisqmin$ for the Centre, as a function of \dmo,
\av\ and metallicity, are presented in the left panels of
Fig.~\ref{chi2_map}. It may be noticed that the best solution is
indeed for $\feh=-0.44$, $\dmo=18.58$, and $\av=0.50$, with a
$\chisqmin=0.62$. Such a small \chisqmin\ is already an indication of
an excellent fit to the observational data.

This best-fitting solution and map of residuals are also presented in
the Hess diagrams of Fig.~\ref{data_model}. Finally, the best-fitting
solution for the cluster Centre is in the left panel of Fig.~\ref{sfr}.

\subsubsection{Evaluating the errors}
\label{sec_erroranalysis}

To evaluate the errors for all involved parameters, the first step is
to find the correspondence between the $\chisqmin$ value for each
model and its significance (or confidence) level, $\alpha$.  This
correspondence was estimated simulating 100 synthetic CMDs generated
with a number of stars equal to the observed CMD, using the
best-fitting SFR$(t)$ and its parameters as the input for the
simulations. So, after recovering the SFH for this sample of synthetic
CMDs, it was possible to build the $\chisqmin$ distribution and to
establish the relation between the $\chisqmin$ difference above the
minimum and $\alpha$.

In the $\chisqmin$ maps of Fig.~\ref{chi2_map}, we superimposed the
68~\% and 95~\% significance levels for all the solutions for the
centre. Only for the $\feh=-0.44$ map we find ample areas of the \av\
versus \dmo\ diagram with solutions within the 68~\% significance
level of the best solution. Based on this figure, we determine
$\dmo=18.58\pm0.07$ and $\av=0.50\pm0.05$ for the cluster Centre (with
random errors at the 68~\% significance level).
  
The left panel of Fig.~\ref{sfr} shows the SFR$(t)$ for the cluster
Centre together with error bars. The most basic feature in this plot
is that the SFR$(t)$ is clearly non-null for three age bins, spanning
the $\log(t/{\rm yr})$ interval from 9.05 to 9.2 (ages from 1.12 to
1.58 Gyr). Note that this result is not only valid for the best
fitting model, but also across the entire 68~\% significance level
area of the \av\ versus \dmo\ diagram. Moreover, it is non-null even
in the case we adopt more restrictive limits for the random errors,
i.e.\ if we plot the random errors for the 95~\% significance level.

\begin{figure}
\resizebox{\hsize}{!}{\includegraphics{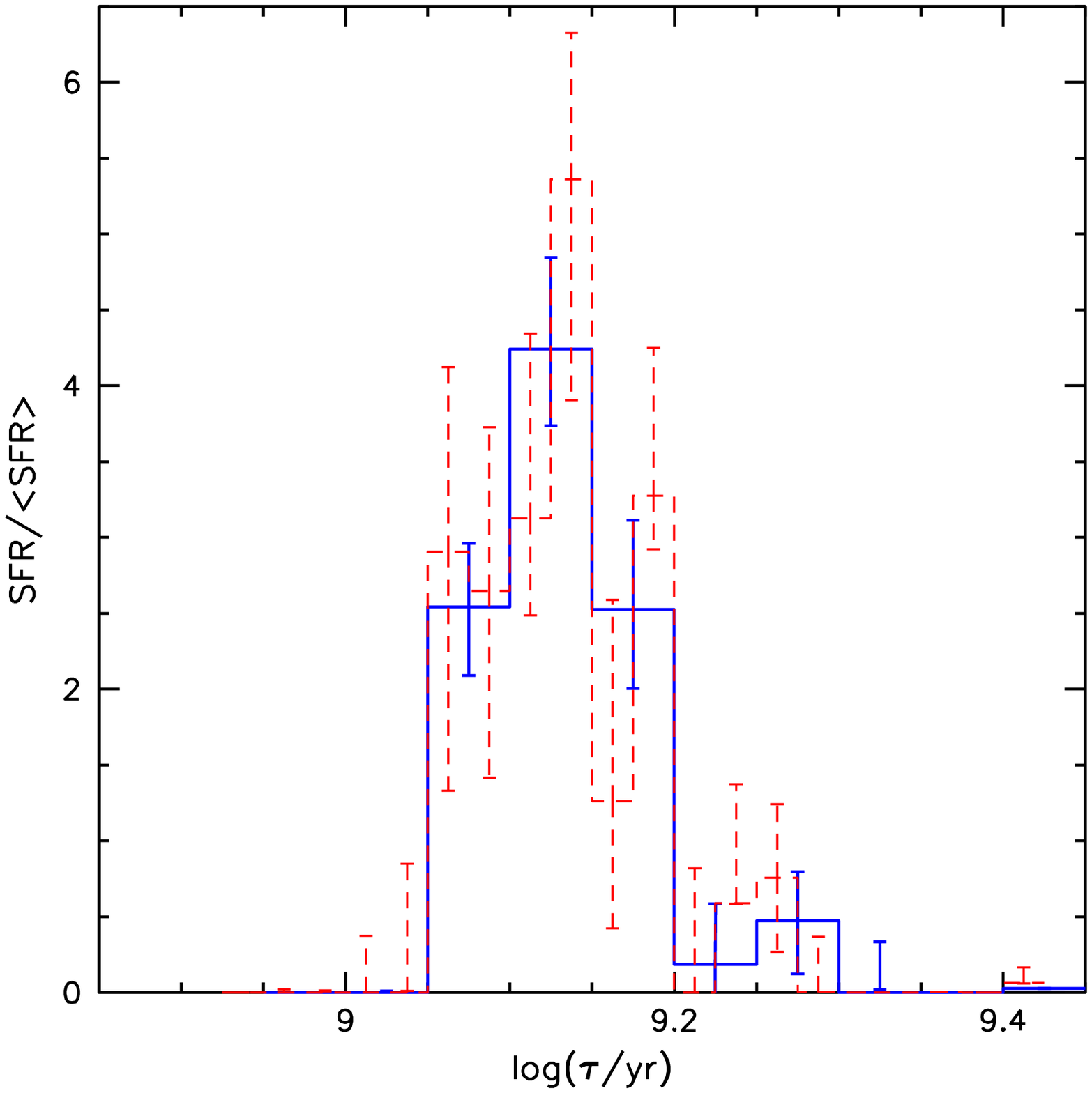}}
\caption{Comparison between the SFR$(t)$ for two best-fitting 
solutions for the cluster Centre, obtained with different age
resolutions: with $\Delta\log t=0.050$~dex (blue continuous line) and
with $\Delta\log t=0.025$~dex (red dashed line).}
\label{fig_fine_SFH}
\end{figure}

Then, one may wonder how the solution for the Centre changes if we
adopt a better age resolution in the SFH-recovery. This is shown in
Fig.~\ref{fig_fine_SFH}, where we compare the solution for
$\Delta\log({\rm age})=0.05$~dex with the one obtained with the same
data and methods, but for an age resolution of $\Delta\log({\rm
age})=0.025$~dex. As we see, within the error bars the two solutions
are essentially the same. The finer resolution in age is compensated
by an increase in random errors.

Therefore, we find evidence that in the NGC~1751 Centre the SFR$(t)$
has lasted for a timespan of 460~Myr. This is about twice longer than
the $\sim200$~Myr estimated by \citet{Milone_etal08} for the same
cluster, based on a simple comparison with the MSTO locations of
different isochrones. 

\subsection{The SFH for the cluster Ring}


We have performed the same experiments of SFH-recovery separately for
the Ring, as illustrated in the $\chisqmin$ maps at the central and
right panels of Fig.~\ref{chi2_map}. In these cases, the levels of
$\chisqmin$ are significantly higher than for the cluster Centre. This
result may seem surprising, considering that the Ring has a lower
level of crowding and hence deeper photometry than the Centre.
We consider as unlikely that these higher $\chisqmin$ for the Ring
could be simply caused by its higher level of contamination from the
LMC field, since this field is very well modeled anyway, and fully
taken into account in the SFH recovery.

Instead, the main reason for the worst fits could be on the presence
of a differential reddening of about $\ebv\sim0.10$~mag within the ACS
field, as found by \citet{Milone_etal08} and Goudfrooij et al. (in
preparation).

\begin{figure*}
\resizebox{0.4\hsize}{!}{\includegraphics{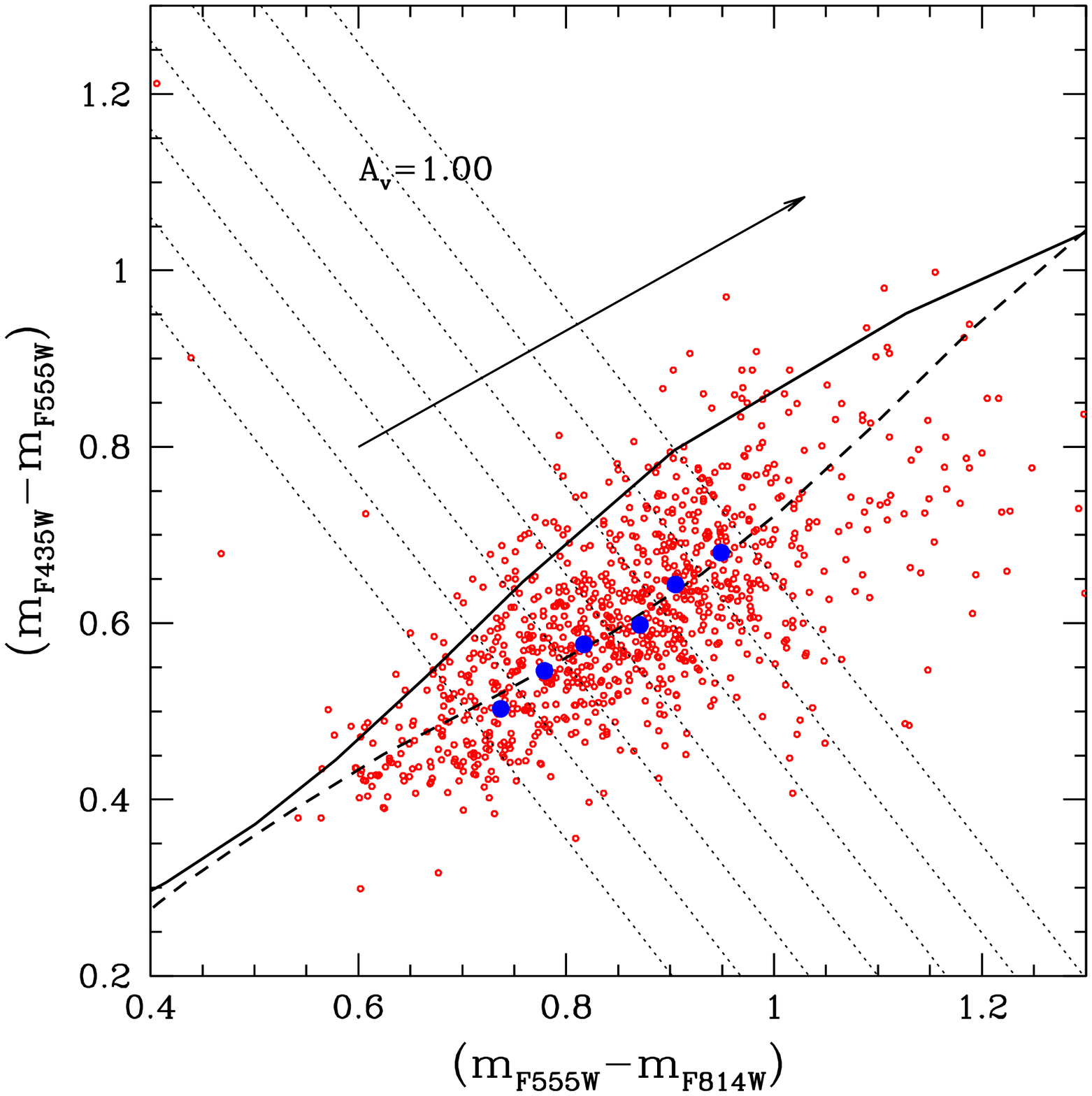}}
\resizebox{0.4\hsize}{!}{\includegraphics{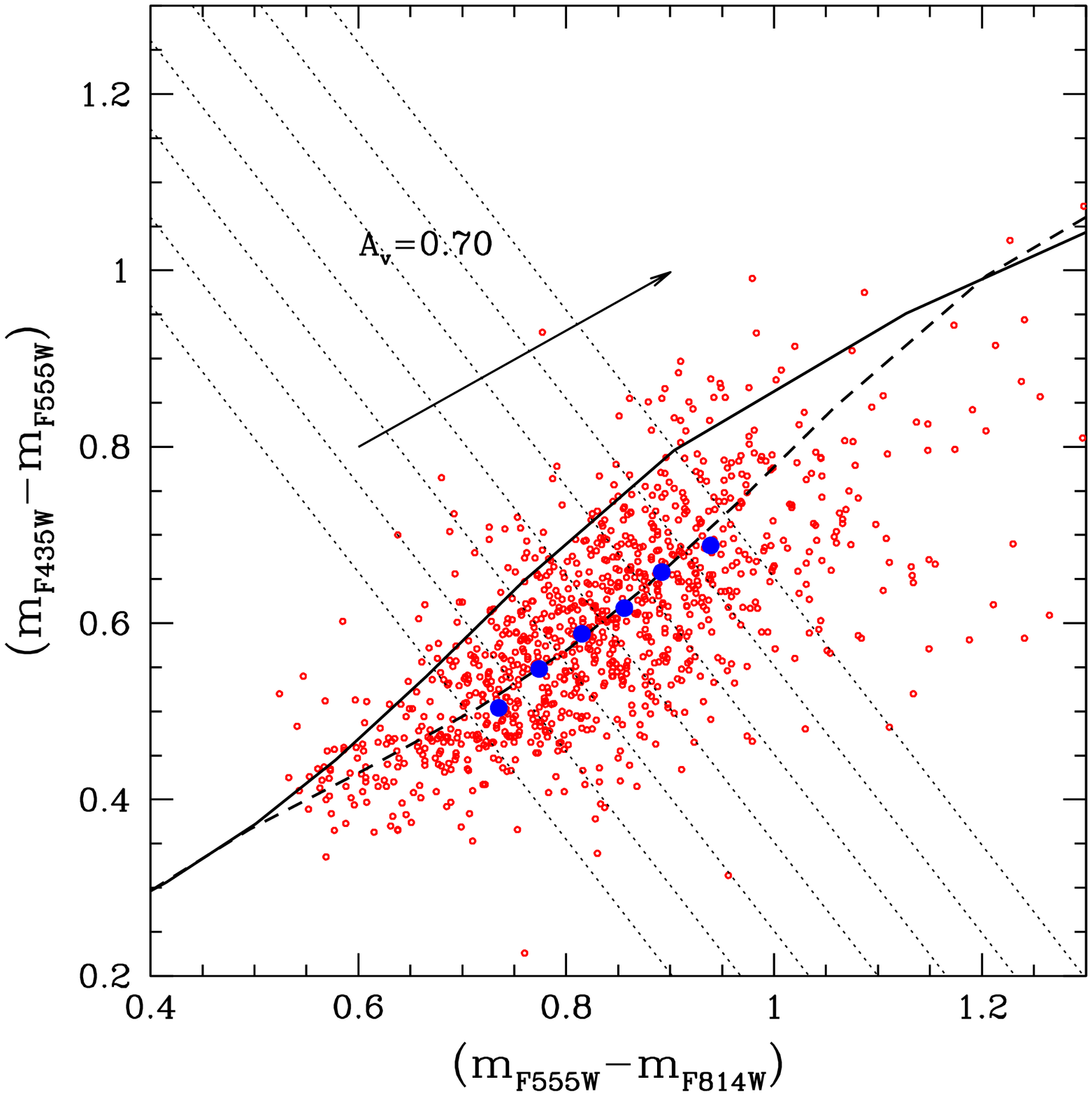}}
\caption{Colour-colour diagrams for the northern (left panel) and 
southern (right panel) half-sections of the Ring, illustrating the
method used to determine the presence of differential reddening. Only
the data for the lower main sequence, with F814W between 21.5 and
23.5, is plotted (red dots). The blue dots mark the median position of
the observed stars in small colour bins (dotted lines) perpendicular
to the reddened main sequence.  The continuous dark line is an
unreddened model zero-age main sequence.  The dashed line shows the
position of the reddened main sequence that best fit these blue
dots. Although the reddening vector (arrow) runs almost parallel to
the main sequence, it is clear that the median position of the points
in this plot can be used to derive the mean reddening in a field. We
do it differentially, determining that between the northern and
southern sections of the Ring there is a difference in $A_V$ of
0.30~mag, which corresponds to 0.10~mag in \ebv.  }
\label{fig_diffreddening}
\end{figure*}

To test for the presence of differential reddening we have followed a
similar procedure as described by \citet{Milone_etal08}, following the
position of fiducial lines in the F435W$\,-\,$F555W
vs. F555W$\,-\,$F814W diagram, and as illustrated in
Fig.~\ref{fig_diffreddening}. So by means of the relative shifts in
the fiducial lines in this colour-colour diagram along the reddening
arrow we have found an extra reddening in the Ring region in relation
to the Centre, with a magnitude in $E_{\rm F555W\,-\,F814W}$ similar
to the one presented by \citet{Milone_etal08}. This extra reddening
occurs prevalently in the bottom and upper extremities of the Ring in
Fig.~\ref{fig_areas}.

Notice also that the extinction values for the Ring, found during the
experiments of SFH-recovery are systematically higher than the ones
found for the Centre region (see Fig.~\ref{chi2_map}), which
independently confirms the presence of differential reddening detected
by us and \citet{Milone_etal08} using fiducial lines.

Unfortunately the recovered best-fitting \dmo\ and \feh\ values for
the Ring are obviously spurious, since they are not consistent with
the ones found for the Centre -- which is a such small region that one
can consider it free from differential reddening.
Considering the high quality of the SFH-recovery for the cluster
Centre (as demonstrated by the much smaller $\chisqmin$), we assume
that the correct distance and reddening of the Ring are the same ones
as for the Centre, namely $\dmo=18.58$ and $\av=0.50$. The right panel
of Fig.~\ref{sfr} shows the Ring SFH for these parameters.

It is interesting to note that the Ring SFR$(t)$, considering just the
random errors, seems to be slightly more spread in age than the one
for the Centre. Indeed, the SFR$(t)$ is found to be non-null in an
additional, older age bin, as compared to the cluster Centre. However,
when one considers the systematics errors in this bin, it is clear
that this result is not solid. It becomes then impossible to take any
conclusion in relation to a possible dependence of the SFR$(t)$ with
the cluster radius.

\section{Concluding remarks}
\label{conclu}

In this work, we confirm that NGC~1751 hosts a dual red clump, similar
to the one observed in the SMC cluster NGC~419
\citep{Girardi_etal09}. Then, we perform SFH-recovery via the
classical method of CMD reconstruction with the sum of single-burst
stellar populations. The basic result is that in the NGC~1751 Centre
star formation is found to last for a timespan of 460~Myr. The same
result is consistently found irrespective of the method used to take
field contamination into account, of the age resolution adopted, and
for a significant region of the \av\ versus \dmo\ plane. The age
resolution of the method is at least 3 times smaller than this
interval. In addition, the best-fitting model is indeed a quite good
representation of the observed data. On the other hand, the CMD for
the cluster Centre is so obviously sharp and clean from the LMC field,
that it is hard to imagine that some important effect has not been
properly considered in our analysis.

Our results contrast with those obtained by \citet{Milone_etal08}, who
identify just two episodes of star formation separated by 200~Myr in
this cluster, using the method of isochrone fitting.

For the cluster Ring, the results indicate a SFR timespan of at least
460~Myr, with a significantly lower quality of the results, as
indicated by the larger \chisqmin\ and by the best fitting model being
found for a distance different from the cluster Centre. These failures
probably depend on the presence of significant differential reddening
across the Ring. So, we prefer not to take any conclusion from this
region. Unfortunately, our method finishes in not providing any
indication about possible variations in the spatial distribution of
the stars with different ages, which would be important for the goals
of understanding the mechanism of extended star formation
\citep[see][]{Goudfrooij_etal09}.

Together with our previous findings for NGC~419
\citep{Rubele_etal10}, the results for the NGC~1751 Centre argue 
in favour of multiple star formation episodes (or continued star
formation) being at the origin of the MMSTOs in Magellanic Cloud
clusters with ages around 1.5~Gyr. The hypothesis of a spread caused
by the presence of fast rotators among the upper main sequence stars
\citep{BastiandeMink09} is separately discussed by 
\citet{Girardi_etal11}, who conclude that it does not offer a valid
alternative to the conclusions reached in the present paper.

\section*{Acknowledgments}
We thank the anonymous referee for the useful suggestions. VKP is
grateful to Jay Anderson for sharing his ePSF program.

The data presented in this paper were obtained from the Multimission
Archive at the Space Telescope Science Institute (MAST). STScI is
operated by the Association of Universities for Research in Astronomy,
Inc., under NASA contract NAS5-26555.  We thank the support from
INAF/PRIN07 CRA 1.06.10.03, contracts ASI-INAF I/016/07/0 and ASI-INAF
I/009/10/0, and the Brazilian agencies CNPq and FAPESP.
%

%
\label{lastpage}

\begin{thebibliography}{}

\bibitem[\protect\citeauthoryear{{Bastian} \& {de Mink}}{{Bastian} \& {de
  Mink}}{2009}]{BastiandeMink09}
{Bastian} N.,  {de Mink} S.~E.,  2009, \mnras, 398, L11

\bibitem[\protect\citeauthoryear{Bonatto 
\& Bica}{2007}]{BonattoBica07} Bonatto C., Bica E., 2007, MNRAS, 377, 1301 

\bibitem[\protect\citeauthoryear{{Bressan}, {Chiosi} \& {Bertelli}}{{Bressan}
  et~al.}{1981}]{Bressan_etal81}
{Bressan} A.~G.,  {Chiosi} C.,    {Bertelli} G.,  1981, \aap, 102, 25

\bibitem[\protect\citeauthoryear{{Carrera}, {Gallart}, {Hardy}, {Aparicio} \&
  {Zinn}}{{Carrera} et~al.}{2008}]{Carrera_etal08}
{Carrera} R.,  {Gallart} C.,  {Hardy} E.,  {Aparicio} A.,    {Zinn} R.,  2008,
  \aj, 135, 836

\bibitem[\protect\citeauthoryear{{Chabrier}}{{Chabrier}}{2001}]{chabrier01}
{Chabrier} G.,  2001, \apj, 554, 1274

\bibitem[\protect\citeauthoryear{{Cioni}, \& {et al.}}{{Cioni} et~al.}{2011}]{Cioni_etal08}
{Cioni} M., {et al.} 2011, \aap, in press, arXiv:1012.5193

\bibitem[\protect\citeauthoryear{{Dolphin}}{{Dolphin}}{2002}]{Dolphin02}
{Dolphin} A.~E.,  2002, \mnras, 332, 91

\bibitem[\protect\citeauthoryear{{Elson}, {Sigurdsson}, {Davies}, {Hurley} \&
  {Gilmore}}{{Elson} et~al.}{1998a}]{Elson_etal98}
{Elson} R.~A.~W.,  {Sigurdsson} S.,  {Davies} M.,  {Hurley} J.,    {Gilmore}
  G.,  1998a, \mnras, 300, 857

\bibitem[\protect\citeauthoryear{{Elson}, {Sigurdsson}, {Davies}, {Hurley} \&
  {Gilmore}}{{Elson} et~al.}{1998b}]{Elson_etal1998}
{Elson} R.~A.~W.,  {Sigurdsson} S.,  {Davies} M.,  {Hurley} J.,    {Gilmore}
  G.,  1998b, \mnras, 300, 857

\bibitem[\protect\citeauthoryear{{Gallart}, {Freedman}, {Aparicio}, {Bertelli}
  \& {Chiosi}}{{Gallart} et~al.}{1999}]{Gallart_etal99}
{Gallart} C.,  {Freedman} W.~L.,  {Aparicio} A.,  {Bertelli} G.,    {Chiosi}
  C.,  1999, \aj, 118, 2245

\bibitem[\protect\citeauthoryear{{Girardi}}{{Girardi}}{1999}]{Girardi99}
{Girardi} L.,  1999, \mnras, 308, 818

\bibitem[\protect\citeauthoryear{{Girardi}, {Bressan}, {Bertelli} \&
  {Chiosi}}{{Girardi} et~al.}{2000}]{Girardi_etal00}
{Girardi} L.,  {Bressan} A.,  {Bertelli} G.,    {Chiosi} C.,  2000, \aaps, 141,
  371

\bibitem[\protect\citeauthoryear{{Girardi}, {Rubele} \& {Kerber}}{{Girardi}
  et~al.}{2009}]{Girardi_etal09}
{Girardi} L.,  {Rubele} S.,    {Kerber} L.,  2009, \mnras, 394, L74

\bibitem[\protect\citeauthoryear{{Girardi}, {Rubele} \& {Kerber}}{{Girardi}
  et~al.}{2010}]{Girardi_etal09rio}
{Girardi} L.,  {Rubele} S.,    {Kerber} L.,  2010, in {R.~de Grijs \&
  J.~R.~D.~L{\'e}pine} ed., IAU Symposium Vol.~266 of IAU Symposium, {Star
  clusters with dual red clumps}.
pp 320--325

\bibitem[\protect\citeauthoryear{Girardi, Eggenberger, 
\& Miglio}{2011}]{Girardi_etal11} Girardi L., Eggenberger P., Miglio A., 2011, MNRAS, L204 

\bibitem[\protect\citeauthoryear{{Glatt}, {Grebel}, {Sabbi}, {Gallagher},
  {Nota}, {Sirianni}, {Clementini}, {Tosi}, {Harbeck}, {Koch}, {Kayser} \& {Da
  Costa}}{{Glatt} et~al.}{2008}]{Glatt_etal08}
{Glatt} K.,  {Grebel} E.~K.,  {Sabbi} E.,  {Gallagher} J.~S.,  {Nota} A.,
  {Sirianni} M.,  {Clementini} G.,  {Tosi} M.,  {Harbeck} D.,  {Koch} A.,
  {Kayser} A.,    {Da Costa} G.,  2008, \aj, 136, 1703

\bibitem[\protect\citeauthoryear{{Goudfrooij}, {Puzia}, {Kozhurina-Platais} \&
  {Chandar}}{{Goudfrooij} et~al.}{2009}]{Goudfrooij_etal09}
{Goudfrooij} P.,  {Puzia} T.~H.,  {Kozhurina-Platais} V.,    {Chandar} R.,
  2009, \aj, 137, 4988

\bibitem[\protect\citeauthoryear{{Grocholski}, {Cole}, {Sarajedini}, {Geisler}
  \& {Smith}}{{Grocholski} et~al.}{2006}]{Grocholski_etal06}
{Grocholski} A.~J.,  {Cole} A.~A.,  {Sarajedini} A.,  {Geisler} D.,    {Smith}
  V.~V.,  2006, \aj, 132, 1630

\bibitem[\protect\citeauthoryear{{Grocholski}, {Sarajedini}, {Olsen}, {Tiede}
  \& {Mancone}}{{Grocholski} et~al.}{2007}]{Grocholski_etal07}
{Grocholski} A.~J.,  {Sarajedini} A.,  {Olsen} K.~A.~G.,  {Tiede} G.~P.,
  {Mancone} C.~L.,  2007, \aj, 134, 680

\bibitem[\protect\citeauthoryear{{Harris} \& {Zaritsky}}{{Harris} \&
  {Zaritsky}}{2001}]{HZ01}
{Harris} J.,  {Zaritsky} D.,  2001, \apjs, 136, 25

\bibitem[\protect\citeauthoryear{{Harris} \& {Zaritsky}}{{Harris} \&
  {Zaritsky}}{2004}]{HZ04}
{Harris} J.,  {Zaritsky} D.,  2004, \aj, 127, 1531

\bibitem[\protect\citeauthoryear{{Harris} \& {Zaritsky}}{{Harris} \&
  {Zaritsky}}{2009}]{HZ09}
{Harris} J.,  {Zaritsky} D.,  2009, \aj, 138, 1243

\bibitem[\protect\citeauthoryear{{Holtzman}, {Gallagher} III, {Cole}, {Mould},
  {Grillmair}, {Ballester}, {Burrows}, {Clarke}, {Crisp}, {Evans}, {Griffiths},
  {Hester}, {Hoessel}, {Scowen}, {Stapelfeldt}, {Trauger} \&
  {Watson}}{{Holtzman} et~al.}{1999}]{Holtzman_etal99}
{Holtzman} J.~A.,  {Gallagher} III J.~S.,  {Cole} A.~A.,  {Mould} J.~R.,
  {Grillmair} C.~J.,  {Ballester} G.~E.,  {Burrows} C.~J.,  {Clarke} J.~T.,
  {Crisp} D.,  {Evans} R.~W.,  {Griffiths} R.~E.,  {Hester} J.~J.,  {Hoessel}
  J.~G.,  {Scowen} P.~A.,  {Stapelfeldt} K.~R.,  {Trauger} J.~T.,    {Watson}
  A.~M.,  1999, \aj, 118, 2262

\bibitem[\protect\citeauthoryear{{Hu}, {Deng}, {deGrijs}, {Goodwin} \&
  {Liu}}{{Hu} et~al.}{2008}]{Hu_etal08}
{Hu} Y.,  {Deng} L.,  {deGrijs} R.,  {Goodwin} S.~P.,    {Liu} Q.,  2008, ArXiv
  e-prints, 801

\bibitem[\protect\citeauthoryear{{Javiel}, {Santiago} \& {Kerber}}{{Javiel}
  et~al.}{2005}]{Javiel_etal05}
{Javiel} S.~C.,  {Santiago} B.~X.,    {Kerber} L.~O.,  2005, \aap, 431, 73

\bibitem[\protect\citeauthoryear{{Kerber}, {Girardi}, {Rubele} \&
  {Cioni}}{{Kerber} et~al.}{2009}]{Kerber_etal09}
{Kerber} L.~O.,  {Girardi} L.,  {Rubele} S.,    {Cioni} M.-R.,  2009, \aap,
  499, 697

\bibitem[\protect\citeauthoryear{{Kerber}, {Santiago} \& {Brocato}}{{Kerber}
  et~al.}{2007}]{Kerber_etal07}
{Kerber} L.~O.,  {Santiago} B.~X.,    {Brocato} E.,  2007, \aap, 462, 139

\bibitem[\protect\citeauthoryear{{Mackey} \& {Broby Nielsen}}{{Mackey} \&
  {Broby Nielsen}}{2007}]{Mackey_BrobyNielsen2007}
{Mackey} A.~D.,  {Broby Nielsen} P.,  2007, \mnras, 379, 151

\bibitem[\protect\citeauthoryear{{Mackey}, {Broby Nielsen}, {Ferguson} \&
  {Richardson}}{{Mackey} et~al.}{2008}]{Mackey_etal08}
{Mackey} A.~D.,  {Broby Nielsen} P.,  {Ferguson} A.~M.~N.,    {Richardson}
  J.~C.,  2008, \apjl, 681, L17

\bibitem[\protect\citeauthoryear{{Mackey} \& {Gilmore}}{{Mackey} \&
  {Gilmore}}{2003}]{MG03}
{Mackey} A.~D.,  {Gilmore} G.~F.,  2003, \mnras, 338, 85

\bibitem[\protect\citeauthoryear{{Milone}, {Bedin}, {Piotto} \&
  {Anderson}}{{Milone} et~al.}{2009}]{Milone_etal08}
{Milone} A.~P.,  {Bedin} L.~R.,  {Piotto} G.,    {Anderson} J.,  2009, \aap,
  497, 755

\bibitem[\protect\citeauthoryear{{Mucciarelli}, {Carretta}, {Origlia} \&
  {Ferraro}}{{Mucciarelli} et~al.}{2008}]{Mucciarelli_etal08}
{Mucciarelli} A.,  {Carretta} E.,  {Origlia} L.,    {Ferraro} F.~R.,  2008,
  \aj, 136, 375

\bibitem[\protect\citeauthoryear{{Nikolaev}, {Drake}, {Keller}, {Cook},
  {Dalal}, {Griest}, {Welch} \& {Kanbur}}{{Nikolaev}
  et~al.}{2004}]{Nikolaev_etal04}
{Nikolaev} S.,  {Drake} A.~J.,  {Keller} S.~C.,  {Cook} K.~H.,  {Dalal} N.,
  {Griest} K.,  {Welch} D.~L.,    {Kanbur} S.~M.,  2004, \apj, 601, 260

\bibitem[\protect\citeauthoryear{{No{\"e}l}, {Aparicio}, {Gallart}, {Hidalgo},
  {Costa} \& {M{\'e}ndez}}{{No{\"e}l} et~al.}{2009}]{Noel_etal09}
{No{\"e}l} N.~E.~D.,  {Aparicio} A.,  {Gallart} C.,  {Hidalgo} S.~L.,  {Costa}
  E.,    {M{\'e}ndez} R.~A.,  2009, \apj, 705, 1260

\bibitem[\protect\citeauthoryear{{Olsen}}{{Olsen}}{1999}]{Olsen99}
{Olsen} K.~A.~G.,  1999, \aj, 117, 2244

\bibitem[\protect\citeauthoryear{{Olsen} \& {Salyk}}{{Olsen} \&
  {Salyk}}{2002}]{OlsenSalyk02}
{Olsen} K.~A.~G.,  {Salyk} C.,  2002, \aj, 124, 2045

\bibitem[\protect\citeauthoryear{{Pejcha} \& {Stanek}}{{Pejcha} \&
  {Stanek}}{2009}]{PejchaStanek09}
{Pejcha} O.,  {Stanek} K.~Z.,  2009, \apj, 704, 1730

\bibitem[\protect\citeauthoryear{{Pessev}, {Goudfrooij}, {Puzia} \&
  {Chandar}}{{Pessev} et~al.}{2008}]{Pessev_etal08}
{Pessev} P.~M.,  {Goudfrooij} P.,  {Puzia} T.~H.,    {Chandar} R.,  2008,
  \mnras, 385, 1535

\bibitem[\protect\citeauthoryear{Piatti et al.}{1999}]{Piatti_etal99} 
Piatti A.~E., Geisler D., Bica E., Clari{\'a} J.~J., Santos J.~F.~C., Jr., 
Sarajedini A., Dottori H., 1999, AJ, 118, 2865 

\bibitem[\protect\citeauthoryear{{Riess} \& {Mack}}{{Riess} \&
  {Mack}}{2004}]{RiessMack04}
{Riess} A.,  {Mack} J.,  2004, Technical report, {Time Dependence of ACS WFC
  CTE Corrections for Photometry and Future Predictions}.
Space Telescope Science Institute

\bibitem[\protect\citeauthoryear{{Rubele}, {Kerber} \& {Girardi}}{{Rubele}
  et~al.}{2010}]{Rubele_etal10}
{Rubele} S.,  {Kerber} L.,    {Girardi} L.,  2010, \mnras, 403, 1156

\bibitem[\protect\citeauthoryear{{Smecker-Hane}, {Cole}, {Gallagher} III \&
  {Stetson}}{{Smecker-Hane} et~al.}{2002}]{Smecker-Hane_etal02}
{Smecker-Hane} T.~A.,  {Cole} A.~A.,  {Gallagher} III J.~S.,    {Stetson}
  P.~B.,  2002, \apj, 566, 239

\bibitem[\protect\citeauthoryear{{Subramanian} \& {Subramaniam}}{{Subramanian}
  \& {Subramaniam}}{2010}]{SubramanianSubramaniam10}
{Subramanian} S.,  {Subramaniam} A.,  2010, \aap, 520, 24

\bibitem[\protect\citeauthoryear{{van der Marel} \& {Cioni}}{{van der Marel} \&
  {Cioni}}{2001}]{vdMC01}
{van der Marel} R.~P.,  {Cioni} M.-R.~L.,  2001, \aj, 122, 1807

\bibitem[\protect\citeauthoryear{{Zaritsky}, {Harris}, {Thompson} \&
  {Grebel}}{{Zaritsky} et~al.}{2004}]{Zaritsky_etal04}
{Zaritsky} D.,  {Harris} J.,  {Thompson} I.~B.,    {Grebel} E.~K.,  2004, \aj,
  128, 1606

\end{thebibliography}
\end{document}